\documentclass[aps,prd,superscriptaddress,showpacs,showkeys,floatfix,preprintnumbers,nofootinbib,letterpaper]{revtex4-2}
\usepackage[Export]{adjustbox}
\usepackage{setspace}
\usepackage{verbatim}
\usepackage{tikz}
\usetikzlibrary {arrows.meta}
\usepackage{color}
\usepackage{amssymb}
\usepackage{pgfplots}
\usepackage{amsmath,amssymb}
\usepackage{mathtools}
\usepackage{hyperref}

\usepackage{graphicx}  
\usepackage{dcolumn}   
\usepackage{bm}        
\usepackage{amssymb}   
\usepackage{physics}
\usepackage{mathtools}
\usepackage{ulem}
\usepackage{placeins}

\definecolor{Rocco}{rgb}{0.0, 0.5, 0.0}
\definecolor{Resolved?}{rgb}{0.5, 0.0, 0.5}

\newcommand{\QQbar}{{Q\bar Q}}

\newcommand{\mcD}{{\mathcal D}}

\newcommand{\FTEE}{FTE$^2$ }
\newcommand{\FTEEformat}{FTE$^2$}
\newcommand{\lp}{\left}
\newcommand{\rp}{\right}

\begin{document}
\title{Entanglement Enabled Tomography of Flux Tubes in (2+1)D Yang-Mills Theory}
\author{Rocco Amorosso}
\affiliation{Department of Physics and Astronomy, Stony Brook University, Stony Brook, NY 11794, USA}
\author{Sergey Syritsyn}
\affiliation{Department of Physics and Astronomy, Stony Brook University, Stony Brook, NY 11794, USA}
\author{Raju Venugopalan}
\affiliation{Physics Department, Brookhaven National Laboratory, Upton, NY 11973, USA}
\affiliation{CFNS, Department of Physics and Astronomy, Stony Brook University, Stony Brook, NY 11794, USA}
\affiliation{Higgs Center for Theoretical Physics, The University of Edinburgh, Edinburgh, EH9 3FD, Scotland, UK}

\begin{abstract}
We investigate the entangling properties of the color flux tube between a static quark-antiquark pair in pure gauge Yang-Mills theory. In earlier works, we defined a gauge-invariant flux tube entanglement entropy (\FTEEformat), the excess entanglement entropy of a region of gluon fields that can be attributed to the color flux tube, and demonstrated that it is finite in the continuum limit.
\FTEE was shown to have two contributions, one from the vibrations of the QCD string, and the other from its internal (color) degrees of freedom. 
In this work, we further explore the internal color component in (2+1)D Yang-Mills theory for $SU(N_c)$ gauge groups, varying $2\le N_c\le5$. We  
identify a novel physical scale in the theory, the entanglement radius $\xi_0$.  
This radius characterizes the transverse extent of the flux tube that must be completely severed by an entangling region to capture the entanglement entropy of color degrees of freedom. 
The key feature underlying this phenomenon is its topological nature. 
This is revealed through systematic studies of multi-slab entangling regions in which \FTEE changes sharply when boundaries of the slabs completely cross-cut the flux tube.
We find that $\xi_0$ increases approximately linearly with $N_c$ and is independent of both R\'{e}nyi replica number and the inter-quark separation length.
We also study \FTEE as a function of the entangling region's transverse displacement from the static quark pair and observe behavior consistent with a  previously identified intrinsic width  $\lambda$ of the flux tube, with an extracted value in agreement with the inverse mass of the lightest glueball for the gauge groups studied.
\pacs{11.15.Ha}
\end{abstract}

\date{\today}

\maketitle
\tableofcontents

\section{Introduction}
Entanglement entropy characterizes the quantum correlations between a subsystem and its complement. It can quantify both local and nonlocal correlations, counting the effective number of degrees of freedom shared between subsystems.
To compute a state's expectation values in a subregion $V$, one needs only the part of the density matrix specific to $V$, with the quantum fields in its complement $\bar{V}$ traced over. This reduced density matrix takes the form 
\begin{equation}
\label{eqn:rho_reduced}
\hat{\rho}_V\equiv\Tr_{\bar V}\hat{\rho}\,,
\end{equation}
where $\hat{\rho}$ is the full density matrix and $\Tr_{\bar V}$ represents the partial trace over fields in $\bar{V}$.
The reduced density matrix can then be used to quantify the entanglement entropy, with the von Neumann entanglement entropy taking form
\begin{equation}
\label{eqn:VNEE}
   S_E=-\Tr \lp(\hat{\rho}_{V} \ln \hat{\rho}_V\rp)\,.
\end{equation}
Since this is often intractable to compute, one can instead calculate the R\'{e}nyi entanglement entropy of order $q$, a measure of entanglement defined as 
\begin{equation}
\label{eq:RenyiEE}
    S^{(q)}=\frac{1}{1-q}\ln\lp(\Tr\hat{\rho_V}^q\rp)\,.
\end{equation}
In quantum field theories, fields in a spatial region are strongly entangled with those in the complement region; the entanglement entropy diverges with the UV cutoff $\Lambda\to\infty$ (or, on the lattice, in the continuum limit $a\to0$)~\cite{Bombelli:1986rw,Srednicki:1993im,Callan:1994py}.
Despite this, there is still a meaningful finite component of the entanglement entropy that can be extracted in the continuum limit. The entanglement entropy typically takes the form~\cite{Calabrese:2004eu,Buividovich:2008kq,Ryu:2006bv,Ryu:2006ef,Nishioka:2006gr,Klebanov:2007ws}
\begin{equation}
\label{eq:entropyUVfinite}
S=S_{UV}+S_f\,,
\end{equation}
where $S$ denotes the entanglement entropy, $S_{UV}$ its UV-divergent component, and $S_f$ its UV-finite component. The UV-finite component can be studied by computing the entropic $C$-function~\cite{Casini:2006es},  calculated on the lattice originally in Refs.~\cite{Buividovich:2008kq,Buividovich:2008gq}.
Studying the entropic $C$-function gives one access to the derivative of $S_f$ with respect to a parameter, provided $S_{UV}$ does not also depend on that parameter.
The entropic $C$-function has been used in many other lattice studies~\cite{Velytsky:2008sv,Itou:2015cyu,Rabenstein:2018bri,Rindlisbacher:2022bhe,Bulgarelli:2023ofi,Jokela:2023rba,Ebner:2024mee,Amorosso:2023fzt,Cataldi:2023xki,Bulgarelli:2024onj,Bulgarelli:2025ewp}. In studying the entanglement entropy of the color flux tube, we found however that it was possible to extract information not only about the derivative of $S_f$, but also $S_f$ itself in this context.

To do so, we introduced a novel Flux Tube Entanglement Entropy (\FTEEformat) defined as  
\begin{equation}
\label{eqn:RenyiDiff}
\tilde{S}^{(q)}_{\vert Q \bar{Q}} = S^{(q)}_{\vert Q \bar{Q}}-S^{(q)} ,
\end{equation}
where $S^{(q)}_{\vert Q \bar{Q}}$ and $S^{(q)}$ are the R\'{e}nyi entanglement entropies of gluon fields in the same spatial region in the presence and absence of static quark sources, respectively~\cite{Amorosso:2024leg}.
\FTEE defined in this way is numerically calculable for gauge theories on a Euclidean lattice, with static quark sources being represented by Polyakov loops.
\FTEE in (2+1)D $SU(2)$ Yang-Mills theory was shown to be finite and non-zero in the continuum limit, having two independent contributions,
\begin{equation}
\label{eq:conjecture}
\tilde{S}^{(q)}_{\vert Q \bar{Q}} = {\tilde S}_\text{vibrational} + {\tilde S}_\text{internal}\,,
\end{equation}
with ${\tilde S}_\text{vibrational}$ representing the entanglement due to the mechanical degrees of freedom of the flux tube and ${\tilde S}_\text{internal}$ representing the entanglement entropy due to color degrees of freedom inside the flux tube.
Motivated by our analytical results in (1+1)D~\cite{Amorosso:2024glf}, where the vibrational entropy is absent,
we conjectured that ${\tilde S}_\text{internal}$ takes form 
\begin{equation}
    {\tilde S}_\text{internal}=\langle F\rangle\ln N_c\,,
\end{equation}
where $\langle F\rangle$ represents the expected number of boundaries crossed by the flux tube.
The expectation value is taken in an effective string model;\footnote{For works on effective string descriptions of the color flux tube, we refer the reader to Refs.~\cite{Polchinski:1991ax,Aharony:2013ipa,Dubovsky:2012sh,Dubovsky:2015zey,Hellerman:2013kba,Hellerman:2014cba,Dubovsky:2013gi}} an example is Luscher's thin string Hamiltonian, which gives Gaussian transverse deflections of the flux tube \cite{Luscher:1980iy}.
This conjecture about ${\tilde S}_\text{internal}$, and lattice calculations in (2+1)D gauge theory supporting it (discussed briefly previously in Ref.~\cite{Amorosso:2025tgg}), constitutes one of the main topics of this paper.
For instance, we study $F>2$ by inserting multiple slabs between quark and antiquark. 
In addition, we also perform studies for gauge groups with $N_c > 2$ exploring the behavior of \FTEE when varying the number of colors and boundary crossings.

This conjecture makes intuitive sense in the strong-coupling limit,
and is best illustrated by \FTEE in (1+1)D.
As we have demonstrated analytically, \FTEE in (1+1)D Yang-Mills takes form
\begin{equation}
    \tilde{S}_{(1+1)YM}=F\ln N_c\,,
\end{equation}
where $F$ is the number of $V/\bar V$ boundaries between the quark $Q$ and antiquark $\bar Q$~\cite{Amorosso:2024glf}.
This result can also be understood in the representation flux basis: Without sources, the (1+1)D ground state is a product of trivial-representation flux states $\vert0\rangle$ on all links and has zero entanglement. With fundamental sources, links between them must carry fundamental-representation flux, and gauge invariance requires the reduced density matrix to carry color indices at any boundary cutting this flux, contributing $\ln N_c$ \FTEE per boundary crossing.
This is discussed in detail in Appendix \ref{sec:AppendixFlux}.

This logic extends to the strong-coupling limit in higher dimensional space.
When we go beyond one spatial dimension, we add a magnetic term to the Hamiltonian, which takes the form
\begin{equation}
\label{eqn:mag_term}
    H_{mag}=f(g)\sum\limits_\Box \text{Tr}(U_\Box+U^\dagger_\Box)\,.
\end{equation}
The prefactor $f(g)$ is inversely proportional to the coupling $g$ with the power dependent on the number of dimensions, $U_\Box$ is a fundamental-representation plaquette matrix 
\begin{equation}
U_\Box=U_{(\vec{x},\vec{x}+\hat{\mu})}U_{(\vec{x}+\hat{\mu},\vec{x}+\hat{\mu}+
\hat{\nu})}U^\dagger_{(\vec{x}+\hat{\nu},\vec{x}+\hat{\mu}+\hat{\nu})}U^\dagger_{(\vec{x},\vec{x}+\hat{\nu})}\,,
\end{equation}
for spatial point $\vec{x}$ and unit directions $\hat{\mu},\hat{\nu}$, and the trace is taken over the color indices.
The magnetic term can mix different flux sectors,\footnote{The magnetic term can be expressed through the magnetic $\hat{R}_{\alpha\beta}$ operators (discussed further in Appendix \ref{sec:AppendixFlux}) that change the representation flux of gauge links.
In the presence of the magnetic term, the ground state is no longer the product state of $\vert0\rangle$ states on links; this state is not an eigenstate of a Hamiltonian carrying magnetic operators.} so that even in the ground state, links may carry higher-representation flux with higher probability as $f(g)$ increases.
In the strong-coupling limit, where $f(g)$ is small and the magnetic term is suppressed, the Hamiltonian looks similar to the (1+1)-dimensional Hamiltonian. The vacuum is composed mostly of links in the $\vert0\rangle$ state, and there is little entanglement as a result.
Adding a fundamental source and sink introduces flux flowing through the links on the shortest path between them, and any boundary intersecting this flux path contributes $\ln N_c$ to the entanglement entropy, for a total of  $F\ln N_c$.

At weak bare coupling (approaching the continuum limit), this simple picture is not valid.
The entanglement entropy of a region $V$ for a general state in lattice gauge theory takes the form~\cite{Donnelly:2011hn}
\begin{equation}
 \label{eqn:entent_flux_rep}
S=H(p(R_\partial))+\sum\limits_{l\in\partial V}\langle\ln dimR_l\rangle+\langle S(\rho_V({R_\partial}))\rangle.
 \end{equation}
In this equation, $R_\partial$ are the flux values (in the group-representation space) on the links on the boundary $\partial V$, $p(R_\partial)$ is their joint probability distribution for the ground state, and $H(p(R_\partial))$ is the Shannon entropy of this distribution.
The dimension of each color representation $R_l$ carried by each link $l$ on the boundary contributes to the entanglement entropy in the second term of Eq.~(\ref{eqn:entent_flux_rep}).
For each assignment of fluxes to links on the boundary, there is a corresponding reduced density matrix $\rho_V(R_\partial)$ and its Von Neumann entropy $S(\rho_V(R_\partial))$.\footnote{Each assignment $R_\partial$ represents a superselection sector, which split the density matrix into block-diagonal form. For more details regarding superselection sectors defining entanglement entropy in gauge theory we refer the reader to Refs.~\cite{Casini:2013rba,Ghosh:2015iwa,Buividovich:2008gq,Donnelly:2011hn,Casini:2013rba,Radicevic:2014kqa,Aoki:2015bsa,Soni:2015yga,Agarwal:2016cir,Lin:2018bud}.}
The averages in the second and third terms are taken with respect to the distribution $p(R_\partial)$.

In the strong-coupling limit, the second term of Eq.~(\ref{eqn:entent_flux_rep}) is zero in the vacuum and $F\ln N_c$ in the presence of static quarks, while all other terms are zero in both cases. 
This is not true in the weak-coupling case.
Towards the continuum limit, $f(g)$ increases and the magnetic term~(\ref{eqn:mag_term}) becomes increasingly relevant.
Gauge links are no longer in an eigenstate of the electric field operator, and instead carry a superposition of different representation labels.

However, our lattice studies~\cite{Amorosso:2024leg,Amorosso:2025tgg} confirm that \FTEE is still $F\ln N_c$ even at weak coupling; it is therefore not immediately obvious from which term of Eq.~(\ref{eqn:entent_flux_rep}) the excess entanglement comes in the continuum limit.
The results of our work here bring some clarity to this issue.
We observe  that the flux tube is a vibrating object with a finite entangling width, which we refer to as its entanglement radius.
This radius has to be fully severed by entangling region $V$
to generate a contribution $\propto\ln N_c$ to \FTEEformat.
Thus, if the $\bar{V}$ portion of the flux tube is split into $G+1$ completely disconnected fragments, the color degrees of freedom will generate $2G\ln N_c$ \FTEEformat.
However, if the flux tube is only partially severed, the color degrees of freedom do not produce any \FTEEformat.
This topological behavior is unlikely to come from the first two terms of Eq.~(\ref{eqn:entent_flux_rep}), as both depend locally on the boundary flux distribution,
and should vary smoothly as the flux tube is gradually cross-cut by an artificial $V/\bar V$ boundary.

This leaves the third term of Eq.~(\ref{eqn:entent_flux_rep}) as the likely source of the $F\ln N_c$ excess entanglement. 
This $\langle S(\rho_V({R_\partial}))\rangle$ term is referred to as the distillable entanglement entropy~\cite{Lin:2018bud}, which can be converted into Bell pairs using local operations and classical communication~\cite{Ghosh:2015iwa,Soni:2015yga}.
Unlike the other two terms in Eq.~(\ref{eqn:entent_flux_rep}), the distillable entanglement entropy is unambiguously dependent on bulk dynamics, not boundary edge modes.
This is in stark contrast to the strong coupling limit discussed above, where \FTEE comes from boundary representation flux degrees of freedom.
It further suggests that \FTEE can be used as a tool to probe quantum information about the internal structure of the flux tube and its dynamics, and not just the dimension of the representation flux carried by it.

It is natural then to ask whether the entanglement entropy can distinguish between different mechanisms of confinement and effective models of the flux tube.
This has been attempted using other observables in similar contexts~\cite{Cardaci:2010tb,Cea:2017ocq,Baker:2025mec,Takahashi:2024vff}.
One common model of confinement is the dual superconductor model of the Yang-Mills vacuum~\cite{Mandelstam:1974pi,tHooft:1977nqb}.
In a superconductor, magnetic flux is expelled via the Meissner effect, and if a magnetic field is forced through a type-II superconductor, the field lines are compressed tightly into thin tubes.
Analogously, the dual superconductor model posits that the QCD vacuum expels chromoelectric flux through the dual Meissner effect.
Inserting quark sources necessitates that the electric flux runs from source to sink, with chromoelectric field lines dynamically compressed into a narrow flux tube, which results in the quark-antiquark potential growing linearly with their separation.
The dual superconductor model has two scales: a penetration depth $\lambda$ and a coherence length $\kappa$ that together parameterize the core of the flux tube.
With this motivation, the Clem formula~\cite{Clem:1975ohd}, created to model the behavior of type-II superconductors, was adapted to the context of non-Abelian gauge theories in Refs.~\cite{Cea:2012qw,Verzichelli:2025cqc,intrinsic:inprep}.
In particular, in Refs.~\cite{Verzichelli:2025cqc,intrinsic:inprep}, the correlator of a plaquette with two Polyakov loops was studied on the lattice.
The Polyakov loops acted as static quark sources and the plaquette probed the field strength depending on its transverse displacement $y$ from the straight line connecting the sources as well as the separation $L$ between them.
The difference between the three-point function and the vacuum expectation value of the plaquette was referred to as $\rho(y)$; 
it was modeled using the Clem formula
\begin{equation}
    \rho(y)\propto K_0\left(\frac{\sqrt{y^2+\kappa^2}}{\lambda}\right)\,,
\end{equation}
where $K_0$ is the Bessel function of the second kind.
The flux tube's field strength was found to fit the Clem formula well, where the penetration depth $\lambda$ was found to be constant as a function of inter-quark separation $L$.

The flux tube's internal structure has also been explored in QED$_3$ in terms of its transverse electric field profile.
As noted, effective string theories predict a roughly Gaussian profile with standard deviation proportional to the logarithm of the source/sink separation~\cite{Luscher:1980iy}.
At low energy, QED$_3$ can also be expressed in the Villain form, and its dynamics can be described with a classical soliton solution.
In this soliton solution, the electric field profile has exponential, not Gaussian, behavior at large transverse distance $y$.
Further, the width of the soliton does not depend on the source/sink separation, only on the mass of the dual photon. The observed electric field profile should thus be a convolution of the two distributions, with a Gaussian center and exponentially decaying tails~\cite{Aharony:2024ctf,Caselle:2016mqu}.
In other words, while the effective string theory yields the wavefunction of the center of the flux tube, the soliton solution describes its internal structure.
We note that in Ref.~\cite{Aharony:2024ctf}, the effective action was analyzed for QED$_3$ for scales between the mass gap and the square root of the string tension, a regime in which both the quantum fluctuations identified with the usual effective string behavior and the classical soliton solution are relevant.

Our work complements these studies on the internal structure of the color flux tube in gauge theories from the standpoint of entanglement. Specifically, we performed numerical studies, using \FTEEformat, of the intrinsic structure of the flux tube emerging between fundamental quarks in (2+1)D Yang-Mills for gauge groups ranging from $SU(2)$ to $SU(5)$.
As a result of our studies, we identify two physical scales:
\begin{enumerate}
    \item The \textbf{entanglement radius} $\xi$, a novel physical property of the flux tube, representing a thickness that must be completely severed by a region $V$ for the region to produce nontrivial \FTEE due to color degrees of freedom.
    \item The \textbf{decay rate} $\lambda$, the scale of \FTEEformat's exponential decay at large transverse displacement of region $V$, which is close in value to a previously identified intrinsic width~\cite{Caselle:2012rp,Verzichelli:2025cqc,intrinsic:inprep}.
\end{enumerate}
These parameters and their behavior as a function of gauge group and other factors will be examined throughout this work, providing deeper insight into effective theories of the color flux tube and the nature of confinement.

The structure of the paper is as follows.
In Section \ref{sec:FTE2Review}, we review how flux tube entanglement entropy is calculated on the lattice, giving the details of the replica structure on the lattice as well as the gauge-invariant lattice observables used to construct \FTEEformat.
In Section \ref{sec:EntanglementRadius}, we introduce the entanglement radius through studying $SU(2)$ (2+1)D Yang-Mills theory.
We begin by summarizing the results of our previous papers on flux tube entanglement, focusing on the internal color entropy contribution.
We compare our results to a thin effective string lacking internal structure.
We find that this model fails when the thin effective string is equally likely to either cross into region $V$ or to miss it;
in particular, we find that the computed value of \FTEE is significantly lower while the effective string model predicts the internal entropy to be at its half-maximum.

We next introduce an alternative geometry of region $V$ with multiple disconnected subregions, varying the gaps between them to further test this regime.
Doing so, we see that \FTEE has a sharp discontinuity as the two subregions are brought into contact and the topology of region $V$ changes.
This is contrary to the 1-D string model, which expects no such discontinuity.
We introduce a revised model of a string endowed with an entanglement radius that must be fully severed by region $V$ in order to contribute non-trivial internal entanglement.
The revised model explains the discontinuity and the modified half-maximum, accurately parameterizes \FTEE in previously considered geometries of $V$, and produces consistent estimates of the entanglement radius $\xi$.
We argue that any effective model of the flux tube must have strong dependency on the topology of the flux tube in $V$ and $\bar{V}$ to accurately reflect our new results.
To explore this further, we study a geometry of $V$ with disconnected subregions with varying lateral positions (normal to the quark separation axis) relative to each other. As a result, we obtain another estimate of the entanglement radius, which agrees with the previous two estimates.
Further evidence for the topological model of flux tube entanglement follows from the observation that the entanglement radius is constant both as a function of length and the order of the R\'{e}nyi entropy $q$.

The ramifications of the topological model of flux tube entanglement are studied further by considering $SU(2)$ through $SU(5)$ (2+1)D Yang-Mills.
Motivated by the topological model of \FTEEformat, we begin by reinterpreting our results in Ref.~\cite{Amorosso:2024leg} as a cumulative distribution function of one of the walls of the flux tube.
With increased statistics, we examine the derivative of \FTEE with respect to the location of region $V$, and find that it produces (1) a peak consistent with the entanglement radius, and (2) the exponential decay at scale $\lambda$.
We study $\lambda$ and $\xi$ as a function of $N_c$ and find that while 
$\lambda$ is roughly constant with a value consistent with the inverse glueball mass, $\xi$ grows approximately linearly with $N_c$.
Section \ref{sec:Conclusions} summarizes our results, discusses their implications, and outlines future work.
In Appendix~\ref{sec:AppendixFlux}, we discuss the representation flux basis and rederive our (1+1)D \FTEE results in terms of flux states.
In Appendix~\ref{sec:Appendix}, we provide the Monte Carlo parameters of our lattice studies as well as scaling data for the entanglement radius for all gauge groups.

\section{Calculating \FTEE on the lattice}
\label{sec:FTE2Review}

\subsection{\FTEE from Polyakov Loops}
In this section, we will specify how we calculate flux tube entanglement entropy on the lattice.
We first review the details of the lattice implementation, specifying the replica geometry of our lattice and how \FTEE is calculated from Polyakov-loop correlators.
We then give the parameters of our Monte Carlo simulations, the results of which are discussed in detail in Section \ref{sec:EntanglementRadius}.

We begin, for the sake of completeness, with a brief review of the replica method on the lattice and the observables used to calculate \FTEE in Ref.~\cite{Amorosso:2024leg}, which we also follow in this work. 
We refer the reader to other works for a more comprehensive review of the underlying issues of gauge invariance of entanglement entropy~\cite{Amorosso:2024leg,Casini:2013rba,Ghosh:2015iwa,Buividovich:2008gq,Donnelly:2011hn,Casini:2013rba,Radicevic:2014kqa,Aoki:2015bsa,Soni:2015yga,Agarwal:2016cir,Lin:2018bud}, previous lattice studies using the same replica construction~\cite{Buividovich:2008kq,Buividovich:2008gq,Velytsky:2008sv,Itou:2015cyu,Rabenstein:2018bri,Rindlisbacher:2022bhe,Bulgarelli:2023ofi,Jokela:2023rba,Ebner:2024mee}, previous studies of \FTEEformat~\cite{Amorosso:2023fzt,Amorosso:2024glf,Amorosso:2024leg,Amorosso:2025tgg}, and alternate treatments of boundary points~\cite{Chen:2015kfa,Amorosso:2024glf}.

For gauge fields $U$ at temperature $1/L_t$, the density matrix can be formally expressed as the path integral in Euclidean space-time
\begin{equation}
\label{eq:dmInOut}
    \langle U^{\text{in}}\vert\rho\vert U^{\text{out}}\rangle=\frac{1}{Z}\int\limits_{U^{\text{in}}}^{U^{\text{out}}}\mathcal{D}Ue^{-S[U]}\,,
    \qquad
    Z = \int\limits_{U^\text{in} = U^\text{out}} \mathcal{D} Ue^{-S[U]}\,,
\end{equation}
where $S$ is the action and $U^{\text{in}}$ and $U^{\text{out}}$ represent the gauge fields at Euclidean time $t=0$ and $t=L_t$ respectively.
The normalization of $\rho$ is chosen  such that its trace is equal to one.
By tracing (integrating) over fields in region $\bar{V}$ with $U^{\text{in}}|_{\bar V} = U^{\text{out}}|_{\bar V}$, one arrives at the reduced density matrix
\begin{equation}
\label{eq:rdmInOut}
     \langle U^{\text{in}}_V\vert\rho_V\vert U^{\text{out}}_V\rangle=\frac{1}{Z}\int\limits_{U^{\text{in}}_V}^{U^{\text{out}}_V}\mathcal{D}U_V\int\limits_{U^{\text{in}}_{\bar{V}}=U^{\text{out}}_{\bar{V}}}\mcD U_{\bar{V}}e^{-S[U_V,U_{\bar{V}}]}\,,
\end{equation}
where the path integral is evaluated over fields on a lattice partially periodic in Euclidean time.
The R\'{e}nyi entanglement entropy~(\ref{eq:RenyiEE}) can be calculated numerically from this formal object as follows:
(1) power $\rho^q$ is expressed  by stacking $q$ \emph{independent} replicas of partially-periodic lattices and identifying the out- and in-states of untraced fields $U_V$ on adjacent replicas; (2) the trace $\Tr(\rho^q)$ is expressed by identifying the in- and out-states of fields $U_V$ in $\rho_V^q$ ; (3) the numerical values are calculated employing standard Monte Carlo algorithms of lattice gauge theory.

\begin{figure}[ht!]
  \centering
\includegraphics[width=.4\textwidth]{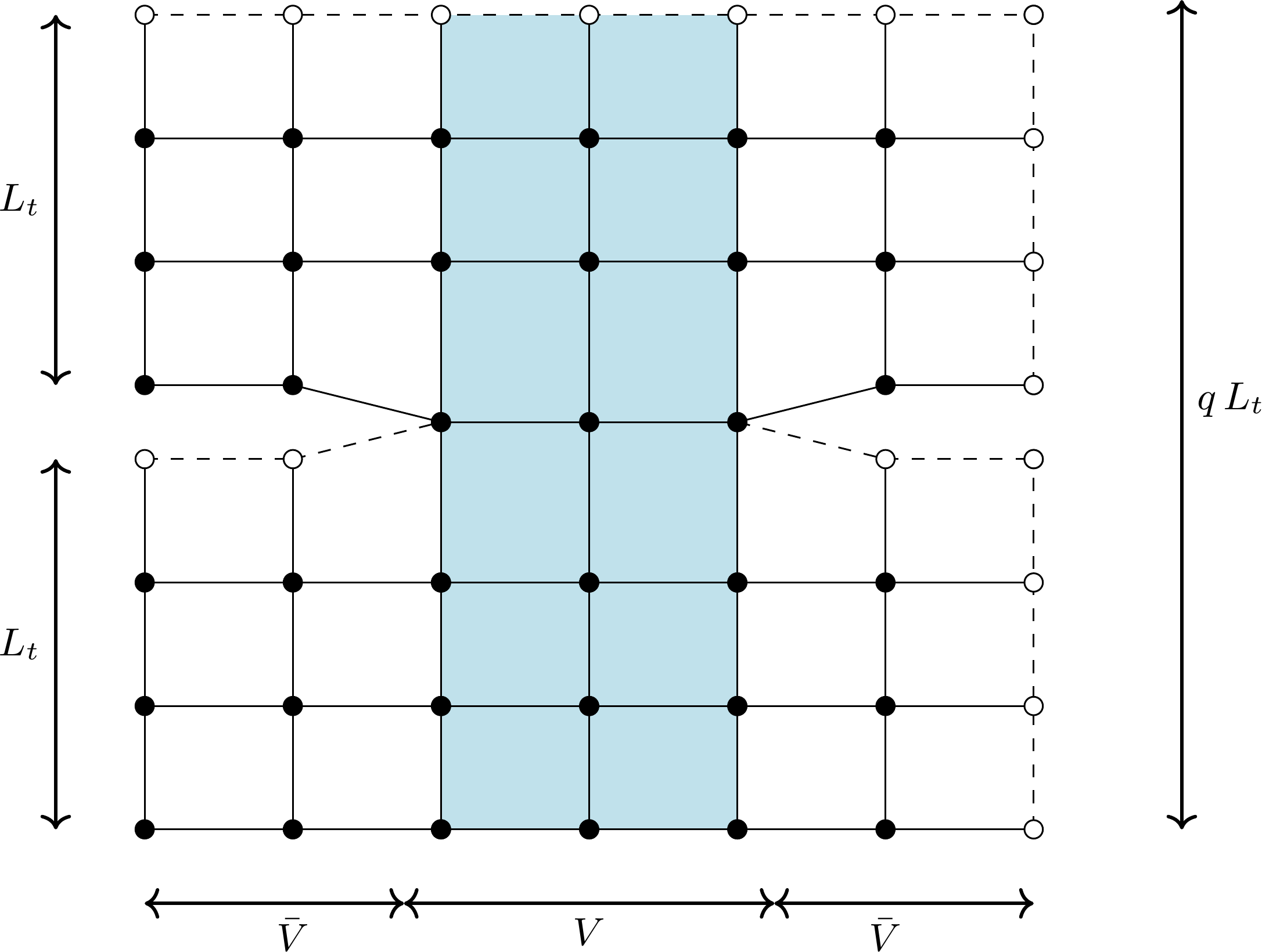}
\caption{\label{fig:pants}
Lattice implementation of $\text{Tr}(\rho_V^q)$ for $q=2$ replicas of an $L_x\times L_t=6\times3$ lattice.
The spatial boundary points have periodicity $L_x$.
The temporal boundary conditions have periodicity $qL_t$ in region $V$ (shaded blue) and $L_t$ in $\bar V$.
Open points and dashed links are images of the solid points and links determined by the spatial and temporal periodic boundary conditions.
To calculate \FTEEformat, one must add temporal Polyakov loops to region $\bar{V}$.
}
\end{figure}

Thus, $\rho_V^q$ corresponds to stacking $q$ replicas labeled by replica number $r\in\{1\dots q\}$, with $U^{\text{in},(r)}_{\bar{V}}$ identified with $U^{\text{out},(r)}_{\bar{V}}$ and $U^{\text{out},(r)}_{V}$ identified with $U^{\text{in},(r+1)}_{V}$ via temporal boundary conditions.
The final trace is then computed by identifying $U^{\text{out},(q)}_{V}$ with $U^{\text{in},(1)}_{V}$ via temporal boundary conditions.
This geometry is shown in Fig.~\ref{fig:pants}.
This reduced density matrix can be used to calculate the entanglement entropy of the vacuum.
To compute \FTEEformat, we follow a very similar idea but also add static quark sources   using temporal Polyakov loops $P_{\vec{x}}=\prod\limits_{t=0}^{L_t-1}U_t(\vec{x},t)$, where $U_t$ represents the gauge link oriented in the $\hat{t}$ direction originating from point $(\vec{x},t)$.
In this work, we will always place quark and antiquark in region $\bar{V}$.
The density matrix of gauge fields in the presence of a single heavy quark at point $\vec{x}$ with initial color state $i$ and final color state $j$ can be expressed as
\begin{equation}
    \label{eq:denmatQuark}
    \langle U^{\text{in}},\vec{x},i\vert\rho\vert U^{\text{out}},\vec{x},j\rangle\propto\langle P_{\vec{x}}\rangle_{ij}\,,
\end{equation}
with the expectation value taken relative to $e^{-S[U]}$.
When a quark (antiquark) source is in $\bar{V}$, to calculate the reduced density matrix, these color states must be traced over.
Explicitly, when inserting a quark-antiquark pair at points $\vec{x}$ and $\vec{y}$, respectively, one has 
\begin{equation}
    \label{eq:denmatQuark2}
    \langle U^{\text{in}}_V\vert\rho_{V\vert Q_{\vec{x}} \bar{Q}_{\vec{y}}} \vert U^{\text{out}}\rangle=\frac{\langle \text{Tr}P_{\vec{x}}\text{Tr}P^\dagger_{\vec{y}}\rangle_{U_{V}^{\text{in}},U_{V}^{\text{out}}}}{\langle \text{Tr}P_{\vec{x}}\text{Tr}P^\dagger_{\vec{y}}\rangle}\,,
\end{equation}
where the numerator is evaluated with periodic temporal boundary conditions in $\bar{V}$, while in $V$ the gauge fields are fixed to $U_V^{\text{in,out}}$ at $t=0$ and $L_t$, respectively.
In order to calculate the numerator explicitly, one would append $\text{Tr}P_{\vec{x}}\text{Tr}P^\dagger_{\vec{y}}$ to the integrand of Eq.~(\ref{eq:rdmInOut}).
The denominator is evaluated with periodic temporal boundary conditions for all space and acts as the normalization factor $Z_{\vert\QQbar}$ for the reduced density matrix in the presence of static quarks, similar to Eq.~(\ref{eq:dmInOut}).
When calculating the trace of the reduced density matrix to the power of $q$, one must stack the replicas with Polyakov loops in $\bar{V}$ as in Fig.~\ref{fig:pants}.
Doing so, $q$ copies of Eq.~(\ref{eq:denmatQuark2}) accumulate, with out-states and in-states of adjacent replicas being identified.
We can then write
\begin{equation}
    \text{Tr}\hat{\rho}_V^q=\frac{Z^{(q)}_{\vert\QQbar}}{(Z_{\vert\QQbar})^q}\,,
\end{equation}
where $Z^{(q)}_{\vert\QQbar}$ is the partition function of the $q$-replica lattice of Fig.~\ref{fig:pants} in the presence of a quark-antiquark pair.
It can be observed that
\begin{equation}
\label{eq:zqqqbar}
\begin{aligned}
Z_{|Q\bar Q} &= Z\,\cdot\,
\left\langle \text{Tr}P_{\vec{x}}\text{Tr}P^\dagger_{\vec{y}}\right\rangle \,,
\\
Z_{|Q\bar Q}^{(q)} &= Z^{(q)}\,\cdot\,
\left\langle \prod\limits_{r=1}^q\text{Tr}P^{(r)}_{\vec{x}}\text{Tr}P^{(r)\dagger}_{\vec{y}}\,\right\rangle \,,
\end{aligned}
\end{equation}
where the first is the usual correlator of two Polyakov loops while the second is
the correlator of $(2q)$ Polyakov loops on a $q$-replica lattice,
and $Z$ and $Z^{(q)}$ are the vacuum partition functions on the usual and $q$-replica lattices, respectively. 
We can then write
\begin{equation}
\label{eq:preFTEEPoly}
        \text{Tr}\hat{\rho}_V^q=\frac{Z^{(q)}_{\vert\QQbar}}{Z^{q}_{\vert\QQbar}}\,=\frac{\langle\prod\limits_{r=1}^q\text{Tr}P^{(r)}_{\vec{x}}\text{Tr}P^{(r)\dagger}_{\vec{y}}\rangle}{(\langle\text{Tr}P_{\vec{x}}\text{Tr}P^\dagger_{\vec{y}}\rangle)^q}\cdot\frac{Z^{(q)}}{Z^q}\,,
\end{equation}
with the expectation value in the numerator evaluated on a $q$-replica lattice and the denominator on an ordinary $L_t$-periodic lattice.
The additional factor $Z^{(q)}/Z^q$ appears due to normalization of Polyakov-loop correlators in Eq.~(\ref{eq:zqqqbar}), and represents the divergent contribution of the vacuum.
When the latter is subtracted, we can express \FTEE using only gauge-invariant lattice observables~\cite{Amorosso:2024leg}
\begin{equation}
\label{eq:FTEEPoly}
\tilde{S}^{(q)}_{\vert\QQbar}\equiv S^{(q)}_{\vert\QQbar} - S^{(q)}
=-\frac{1}{q-1}\left(\ln \frac{Z^{(q)}_{\vert\QQbar}}{Z^{q}_{\vert\QQbar}}-\ln \frac{Z^{(q)}}{Z^q}\right)
=-\frac{1}{q-1}\ln\left( \frac{\langle\prod\limits_{r=1}^q\text{Tr}P_{\vec{x}}^{(r)}\text{Tr}P^{(r)\dagger}_{\vec{y}}\rangle}{(\langle\text{Tr}P_{\vec{x}}P^\dagger_{\vec{y}}\rangle)^q}\right)\,.
\end{equation}

\subsection{Monte Carlo simulation details}
The Monte Carlo calculations in this work were performed in the confining phase of $SU(2)$ through $SU(5)$ (2+1)D Yang-Mills theory.
For the single-slab runs, where region $V$ consists of one rectangular slab, we performed our lattice studies at $T_c/2$.
As discussed in Ref.~\cite{Amorosso:2024leg}, at this temperature, the quark-antiquark system is dominated by its ground state while statistical noise is under control.
Further, this was an economical choice in order to make use of the substantial statistics collected in Ref.~\cite{Amorosso:2024leg}.
For the multi-slab runs, where region $V$ consists of more than one rectangular slab, we performed our lattice studies at $T_c/4$.
Values of $\beta$ were chosen to match the desired temperature using the fit $\beta_c(N_t)=N_c^2(0.380(3)N_t+0.106(11))$ from Ref.~\cite{Billo:1996pu}, then further adjusted to bring the width of the slab closer to $w\sqrt{\sigma_0}\sim0.25$.
This adjustment was made so we could study ensembles across different lattice spacings $a$ and colors $N_c$ with consistent physical parameters.\footnote{Matching the lattice spacing of different gauge groups in string tension units essentially amounts to holding $g_0^2N_c$ constant, where $g_0^2=\frac{2N_c}{\beta}$. This holds especially in the continuum limit ($a\to0$) and can be checked by comparing the bare coupling $g_0$ for different values $N_c$ in Appendix~\ref{sec:Appendix}.}
The lattice spacings $a\sqrt{\sigma_0}$ for our ensembles are listed in Tables~\ref{tab:lattices_runsSU2}--\ref{tab:lattices_runsSU5} in Appendix \ref{sec:Appendix}.
They have either been determined previously from the literature for the given value of $\beta$~\cite{Teper:1998te,Athenodorou:2016ebg}, or determined from the continuum extrapolations listed in Table 18 of Ref.~\cite{Teper:1998te}.
As an additional check for our $N_c>2$ runs, we have calculated the string tension using an ordinary lattice at $T_c/2$, fitting the potential to a linear form plus the Luscher term and finding general agreement with the string tensions from the literature.
The calculated string tensions for all runs are listed in the tables in Appendix \ref{sec:Appendix}.
The re-analyzed $SU(2)$ runs are listed along with the $T_c/2$ string tension determined in Ref.~\cite{Amorosso:2024leg}.

To compute \FTEE on the lattice, we use the standard Wilson plaquette action
\begin{equation}
    S=-\frac{\beta}{N_c}\sum\limits_{p}\text{Re}\text{Tr}U_p\,,
\end{equation}
summing over plaquettes $p$ on the lattice.  
We perform Monte Carlo updates with sweeps consisting of one Kennedy-Pendleton heatbath update~\cite{Kennedy:1985nu} and five overrelaxation updates~\cite{Brown:1987rra}.
The specifics of thermalization and the multilevel algorithm used in this work are identical to those used in Ref.~\cite{Amorosso:2024leg}, to which we refer the reader for further detail.

To suppress correlated errors, the denominator of Eq.~(\ref{eq:FTEEPoly}) is calculated on the $q$-replica lattice sufficiently far from the region $V$.
These values have been checked to match the single-replica result.

\section{The Entanglement Radius}
\label{sec:EntanglementRadius}
In this section, we discuss our results and their implications for our understanding of flux tube structure.
We first review the results of our previous studies of \FTEEformat, investigating the regimes in which a simple 1-D string model fails to explain our results.
This motivates our introduction of the entanglement radius, which we study in detail using a variety of geometries of the entangling region.

\begin{figure}[ht!]
\centering
\includegraphics[width=.35\textwidth,valign=c]{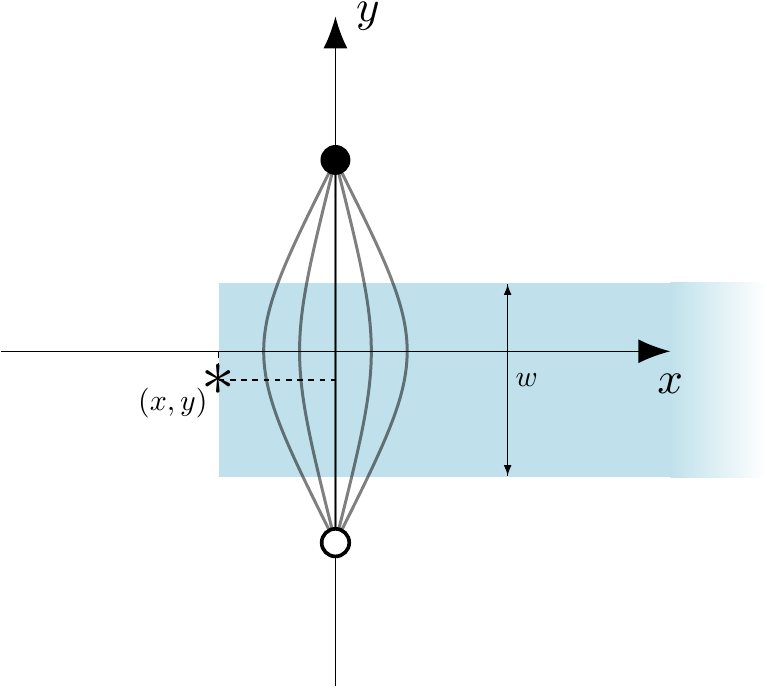}
\caption{\label{fig:refinedhalfslab} 
Flux tube entanglement entropy was computed in Ref.~\cite{Amorosso:2024leg} using the ``half-slab'' geometry depicted above.
Region $V$ (shaded blue) has width $w$ in the $y$-direction and length $L_x/2$ in the $x$-direction.
The flux tube is centered at $(0,0)$ with quark and antiquark separated by distance $L$ in the $y$-direction.}  
\end{figure}

In the original study of \FTEEformat~\cite{Amorosso:2024leg}, the R\'{e}nyi entanglement entropy of the shaded region depicted in Fig.~\ref{fig:refinedhalfslab} was shown to have finite entanglement entropy attributable to the color flux tube in $SU(2)$ Yang-Mills.
Further, \FTEE could be partitioned into a vibrational entropy arising from mechanical vibrations of the flux tube and an internal piece arising from color degrees of freedom inside the flux tube.
For the quark separations studied ($L\sqrt{\sigma_0}\sim1$), we found the entanglement entropy is dominated by the internal  entropy, which contributes to \FTEE an amount proportional to $\ln N_c$, where $N_c$ is the number of colors.
Smaller, but still significant, was the vibrational entropy, taking the form $\frac{1}{4}\ln (L/\delta)$ for a flux tube crossing two boundaries, with $\delta$ the characteristic wavelength of the flux tube, and $L$ its length.

For the purposes of this paper, we will focus on the internal entropy.
The conjectured form of the internal entropy appearing in Ref.~\cite{Amorosso:2024leg} was motivated by the results of \FTEE in (1+1) dimensions.
In the absence of transverse vibrations, a zero-width string is a 1-D system, with flux running between source and sink along its length.
Through Haar integration, we found that \FTEE takes form~\cite{Amorosso:2024glf}
\begin{equation}
\label{eq:FlogNc}
    \tilde{S}=F\ln N_c
\end{equation}
in (1+1) dimensions, with $F$ the number of boundaries of entangling region $V$ crossed by the flux tube, and $N_c$ the number of colors.
This presented a simple picture for generalizing to higher dimensions: a vibrating 1-D string (no transverse structure) free to cross the $V$/$\bar{V}$ boundary an arbitrary number of times, with each boundary crossing being associated with an additional $\ln N_c$ entanglement entropy.
This $\ln N_c$ contribution can be understood as a manifestation of Gauss's law, matching the color flux in $V$ with that of $\bar{V}$ at each intersection between the 1-D string and the boundary.
With this picture in mind, we generalized Eq.~(\ref{eq:FlogNc}) to higher dimensions, writing
\begin{equation}
    \label{eq:expFlogNc}
    S_{int}=\langle F\rangle \ln N_c,
\end{equation}
with $\langle F\rangle$ determined by the geometry of $V$ and the string's mechanical properties, namely its string tension, length, and location of its endpoints (static quark sources).
Using Luscher's thin-string Hamiltonian~\cite{Luscher:1980iy}, one could arrive at a 1-D string with Gaussian distribution of deflection.
This simple model of flux tube entanglement works well when the source and sink are well-separated by region $V$ (the $x\to-\infty$ limit of Fig.~\ref{fig:refinedhalfslab}), with flux strongly preferring to pass through $V$ as opposed to traveling around it.

In Ref.~\cite{Amorosso:2024leg}, we studied the \FTEE of region $V$, a rectangular slab of finite width in the $y$-direction taking up half the $x$-extent of the lattice, with varying $(x,y)$ position relative to the center of the $Q\bar Q$ pair.
This geometry is schematically depicted in Fig.~\ref{fig:refinedhalfslab}.
For large negative $x$, the \FTEE was indeed $\langle F \rangle \ln N_c$.
This was demonstrated for $SU(2)$ with a single half-slab in  
Ref.~\cite{Amorosso:2024leg} and made more robust in Ref.~\cite{Amorosso:2025tgg}, where different numbers of colors and boundary crossings were explored.

\begin{figure}[ht!]
\centering
\includegraphics[width=.49\textwidth,valign=c]{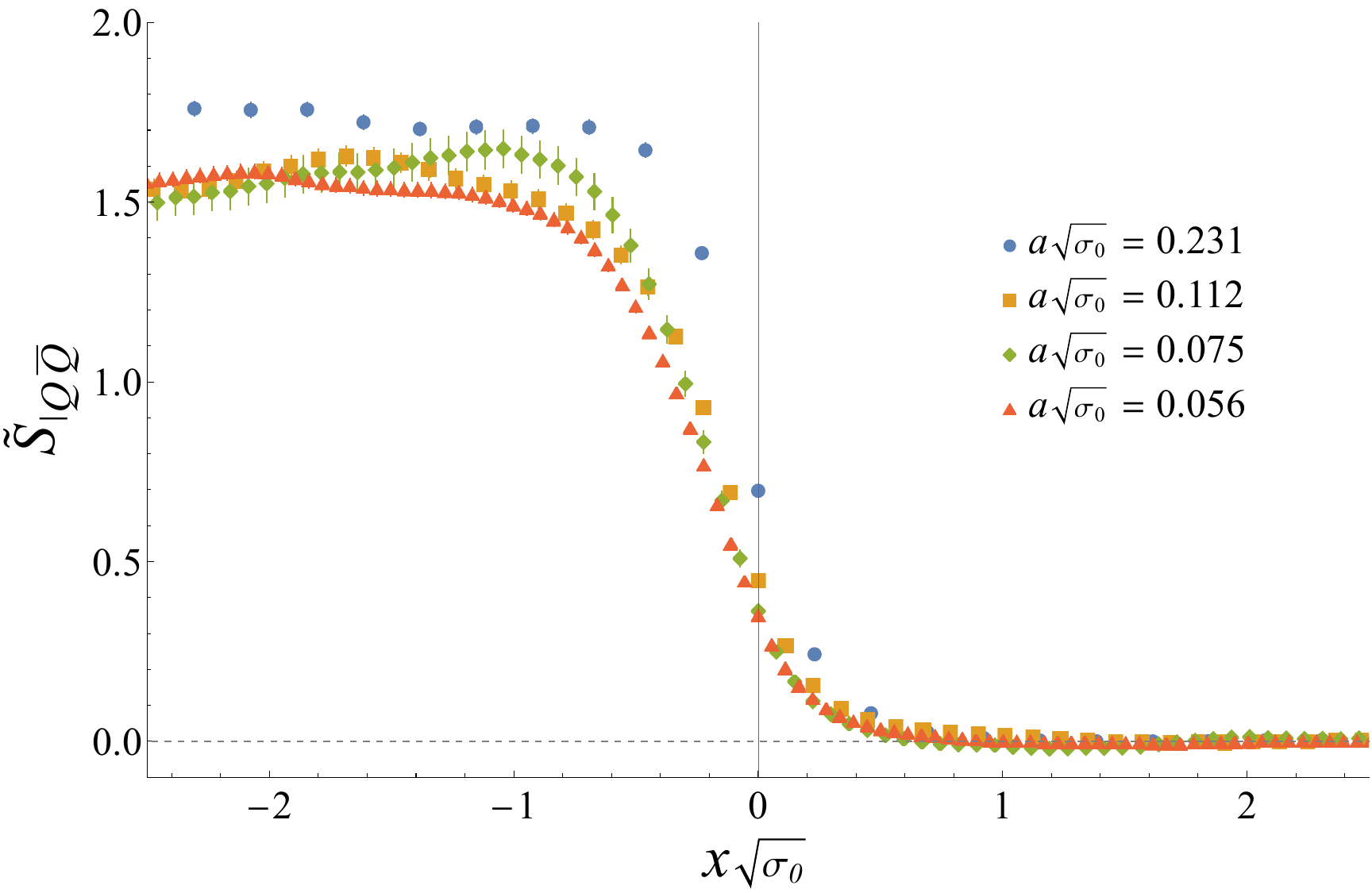}
\includegraphics[width=.49\textwidth,valign=c]{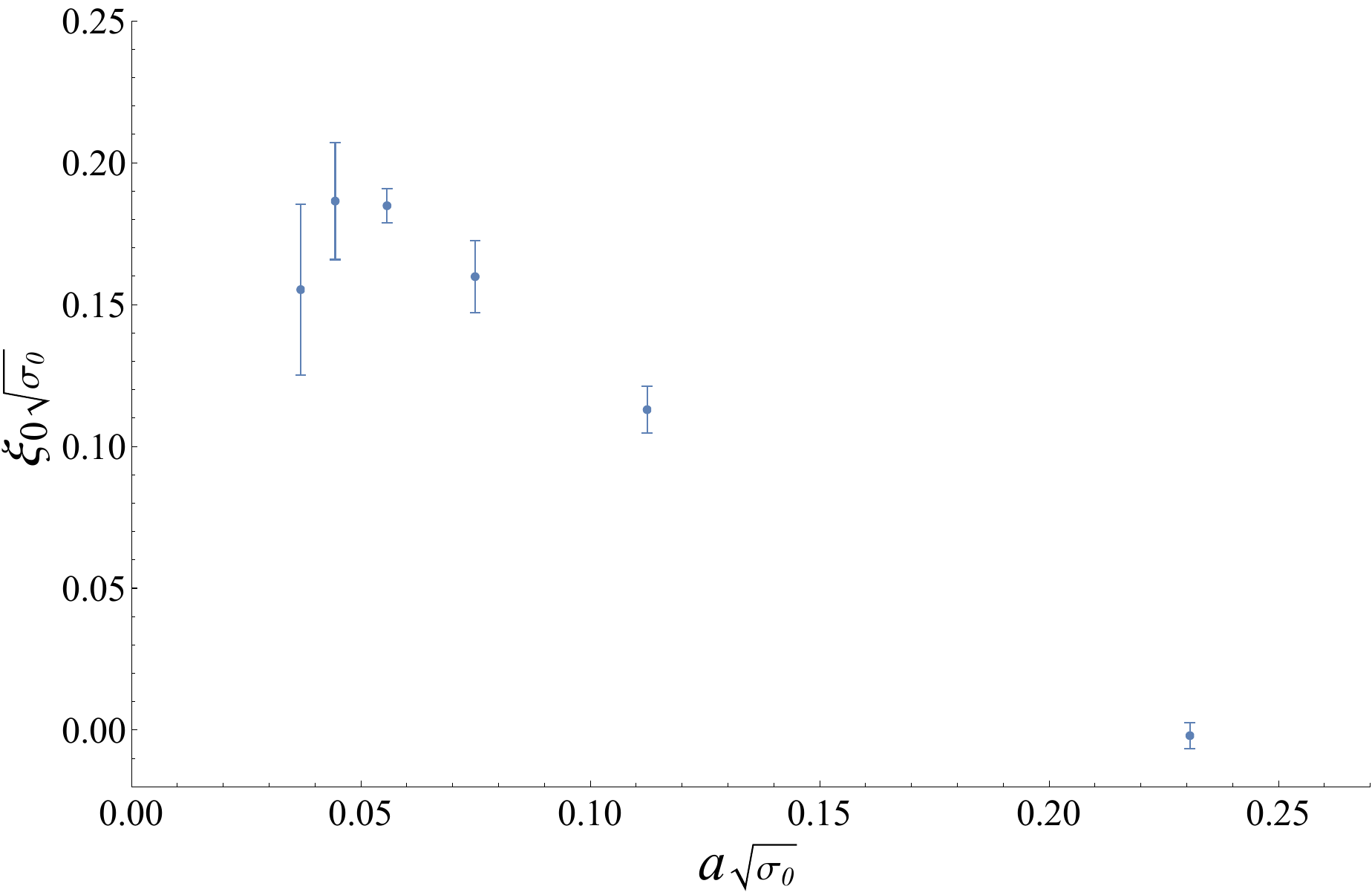}
\caption{\label{fig:ascalingOG} 
(Left) \FTEE of the half-slab geometry in $SU(2)$ (2+1)D Yang-Mills as a function of the lattice spacing $a$ and slab edge position $x$. Results are shown for a fixed inter-quark distance $L\sqrt{\sigma_0}=0.67$, slab width $w\sqrt{\sigma_0}=0.22$, and $y=0$.
(Right) The offset of the half-maximum value of internal \FTEE from $x=0$ (the value $x=-\xi_0$ at which $\tilde{S}=\ln2$) as a function of the lattice spacing. Note that $\xi_0$ converges to a finite value as we approach the continuum limit.}
\end{figure}

While the 1-D string with trivial internal structure was successful at large values of $x$, it had difficulty explaining features of \FTEE at $x\sim0$. 
By moving the half-slab in Fig.~\ref{fig:refinedhalfslab} relative to the quark-antiquark pair, we were able to study \FTEE as a function of $x$ while keeping constant $y=0$.
Contrary to the expectations of the thin string model, \FTEE reached its half-maximum value \emph{not at} $x=0$, when the leftmost edge of $V$ was aligned with the static sources, but at value $x=-\xi_0$ that was finite and non-zero in the continuum limit.
The results of this study are shown in Fig.~\ref{fig:ascalingOG}.
These results are in opposition to the thin string picture, in which the string deflects into $V$ 50\% of the time at $x=0$,\footnote{
As the number of vibrational modes increases, the string would deflect into $V$ more than half the time, but this effect would shift rightward the graph of Fig.~\ref{fig:ascalingOG} (Left), not leftward, as seen in the figure.
}
resulting in $\langle F\rangle\approx1$ and $\tilde S_{|Q\bar Q}\approx \ln 2$ for $N_c=2$.
\begin{figure}[ht!]
\centering
\raisebox{-.5\height}{\includegraphics[width=.23\textwidth]{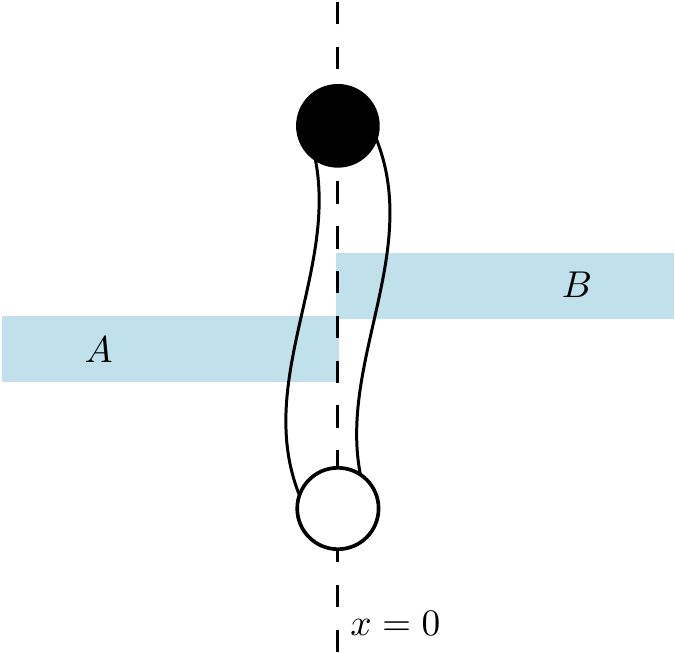}}
\hfill
\raisebox{-.5\height}{\includegraphics[width=.23\textwidth]{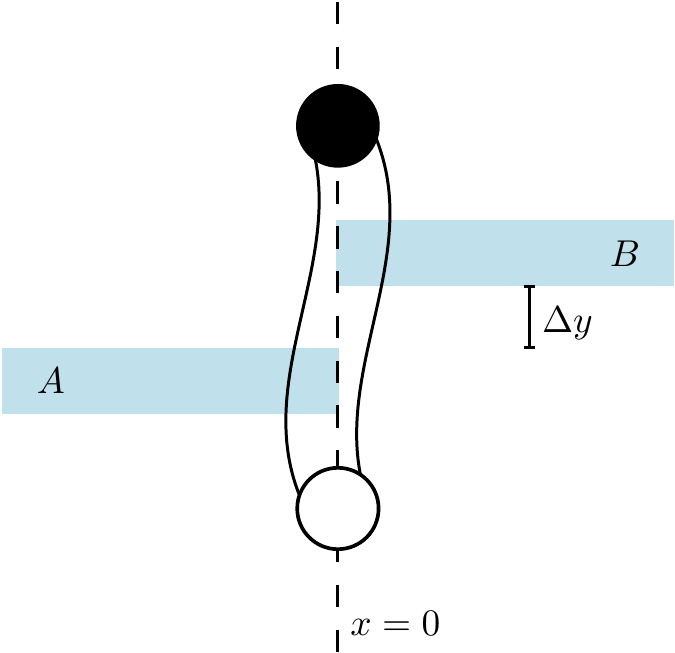}}
\hfill
\raisebox{-.5\height}{\includegraphics[width=.44\textwidth]{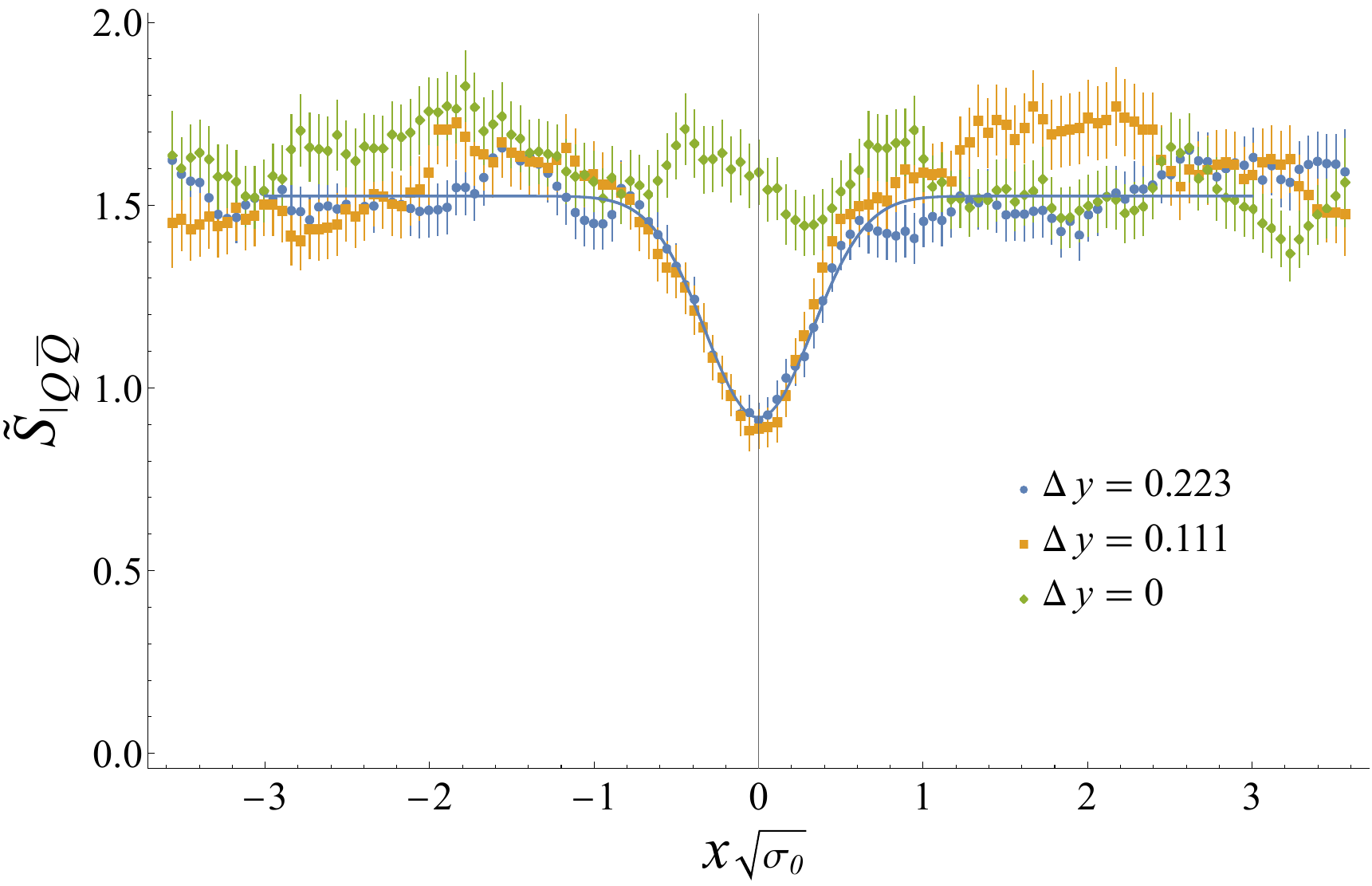}}
\caption{\label{fig:StaggerGeom}
(Left) Schematic of a flux tube exhibiting a full boundary crossing in the staggered two-slab geometry with $\Delta y=0$. Here the flux tube has two disconnected components in $\bar{V}$.
(Center) Schematic of a flux tube exhibiting a partial boundary crossing in the staggered two-slab geometry with $\Delta y>0$. Here the flux tube has one connected component in $\bar{V}$.
(Right) \FTEE of the staggered two-slab geometry in $SU(2)$ Yang-Mills as a function of $x$ for different gap sizes $\Delta y$. A steep drop in \FTEE is seen for non-zero $\Delta y$ at $x=0$. Here $a\sqrt{\sigma_0}=0.056$, the individual slab width is $4a$, and the quark separation $L=20a$. 
The proposed fit form~(\ref{eq:roughEst}) (plus a fitted constant to account for vibrational contributions) is shown with a solid line.
}
\end{figure}
To further study this discrepancy, we examine an alternative geometry, allowing us to study the dependence of \FTEE on the topology of region $V$.
For this study, we place two slabs staggered with no gap in between, as seen in Fig.~\ref{fig:StaggerGeom} (Left), and then separate them by a small distance $\Delta y$ on the $y$-axis, as seen in Fig.~\ref{fig:StaggerGeom} (Center). 
A 1-D vibrating effective string would predict the two-slab geometries in Fig.~\ref{fig:StaggerGeom} (Left) and Fig.~\ref{fig:StaggerGeom} (Center) to produce similar \FTEE values.
Due to the linearity of expectation values, $\langle F \rangle = \langle F_A\rangle + \langle F_B\rangle$, where the number of boundary crossings $F_A$ and $F_B$ for the two disconnected slabs of $V$ have little dependence on their $y$-position, as has been inferred from results of Ref.~\cite{Amorosso:2024leg}.
One can further argue by symmetry that every configuration of a zero-width string that avoids both slabs in Fig.~\ref{fig:StaggerGeom} (Center) and yields $F=0$ is compensated by a configuration reflected in the $x$ direction and yielding $F=4$ with equal probability, resulting in the same $\tilde{S}_{int}\approx2\ln N_c$ color contribution to \FTEE for both the $\Delta y>0$ (Fig.~\ref{fig:StaggerGeom}, Center) and $\Delta y=0$ (Fig.~\ref{fig:StaggerGeom}, Left) geometries.
Thus the vibrating 1-D string model predicts that \FTEE in multiple-slab geometries can have only weak dependence on $\Delta y$. 
However, our results clearly and significantly deviate from this zero-width string expectation.\footnote{One may think the replica geometry of \FTEE introduces an unphysical energy cost to transporting flux across a boundary, artificially suppressing string configurations crossing boundaries. This is contradicted by data (shown later) where the entanglement radius is shown to be unaffected by changing the string length and number of replicas.}
While there is minimal difference between the $\Delta y=4a$ and $\Delta y=2a$ ensembles, \FTEE rises significantly and abruptly when the topology of $V$ changes to become one connected region at $\Delta y=0$.

\begin{figure}[ht!]
\centering
\includegraphics[width=.2\textwidth,valign=c]{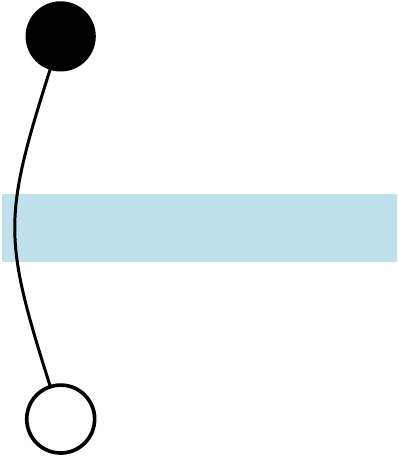}
\hspace{3cm}
\includegraphics[width=.2\textwidth,valign=c]{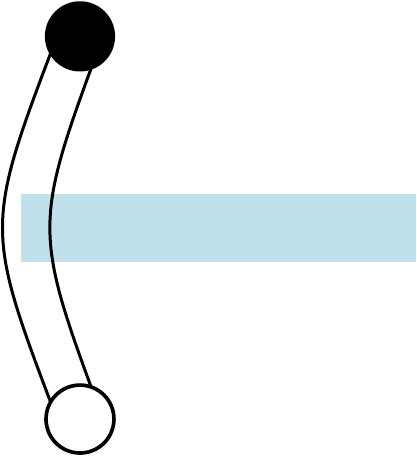}
\caption{\label{fig:BoCr} 
Schematic of two flux tubes with the same deflection (partially) intersecting a region $V$ (shaded blue).
(Left) The flux tube has zero entanglement radius and exhibits a full boundary crossing, being separated into two disconnected components within $\bar{V}$ and resulting in an internal entropy of $2\ln N_c$.
(Right) The flux tube has finite non-zero entanglement radius and exhibits a partial boundary crossing.
The flux tube has one connected component within $\bar{V}$, resulting in zero internal entanglement entropy.
}
\end{figure}

To adequately model flux tube entanglement, we refine our model by endowing our effective string with a finite width and thereby introducing an ambiguity in the formula $\tilde{S}=\langle F\rangle\ln N_c$.
Such a ``thick'' string of finite width can have partial boundary crossings, in which the boundary only partially intersects the string and separates it into two \emph{connected} components (one in $\bar V$ and one in $V$), as shown in Fig.~\ref{fig:BoCr} (Right).
While it is unclear how to treat configurations with partial boundary crossings a priori, by assigning value $F=0$ to partial boundary crossings we can now explain the drastically different results seen in Fig.~\ref{fig:StaggerGeom} (Right).
To examine this hypothesis, we introduce a naive model of flux tube entanglement: (1) the flux tube is represented by a thick vibrating effective string of fixed \textbf{entanglement radius} $\xi_0$;
(2) every string configuration exhibiting a full boundary crossing, where the string is divided into three separate components, is associated with a $\ln N_c$ \FTEE contribution; and
(3) the string has one vibrational mode and has Gaussian-distributed deflection in the $x$-direction.
As there is only one vibrational mode in this naive model, we can approximate the flux tube's deflection as $y$-independent near the center of the flux tube.
If $\chi\sim N(0,w)$ is the Gaussian-distributed transverse deviation of the string center from the straight $Q\bar Q$ line, then the probabilities of the thick string fully crossing half-slabs $A$ and $B$ are
\begin{equation}
\langle p_A\rangle = \int\limits_{-\infty}^{x-\xi_0} p(\chi)\,d\chi \,,
\qquad
\langle p_B\rangle = \int\limits_{x+\xi_0}^{\infty} p(\chi)\,d\chi \,,\\
\end{equation}
where $p(\chi)\propto \exp[-\chi^2/(2w^2)]$. With $\langle F\rangle = 2(p_A+p_B)$, the contribution to the internal entropy is
\begin{equation}
\label{eq:roughEst}
f(x) = 2(p_A + p_B)\ln 2 = 2 \ln 2 \,\left[
1 + \Phi\left(\frac{x-\xi_0}{w}\right) -\Phi\left(\frac{x+\xi_0}{w}\right)\right]\,,
\end{equation}
where $\Phi(t)=\frac{1}{2}\left(1+\text{erf}(t/\sqrt{2})\right)$ is the cumulative normal distribution. 
Above, $w$ represents the width of the Gaussian deflection of the flux tube.
This rough estimate  fits the lattice data surprisingly well, with the 
model curve clearly accounting for the shape of the $\Delta y >0$ geometries at $x\sim0$, as shown in Fig.~\ref{fig:StaggerGeom} (Right).
From the fit, we can extract  $\xi_0$.
Likewise, recall  that the entanglement radius is independently determined by examining the 
offset of \FTEE from $x=0$ in the half-slab geometry\footnote{As the vibrational entropy is indistinguishable from zero at $x\gtrsim0$~\cite{Amorosso:2024leg,Amorosso:2025tgg} and significant (but still small) at large negative $x$, we calculate the point $x=-\xi_0$ at which \FTEE is equal to $\ln N_c$ instead of the half-maximum of \FTEEformat.} and calculating the point $x=-\xi_0$ at which \FTEE is equal to $\ln N_c$.
{\it Both estimates of $\xi_0$ are in excellent agreement}, with $\xi_{0,\text{offset}}\sqrt{\sigma_0}=0.185(6)$ and 
$\xi_{0,\text{fit}}\sqrt{\sigma_0}=0.182(18)$.

Before discussing this simple model further, it is important to stress the topological nature of \FTEEformat.
Looking at the portion of the effective string in $\bar{V}$, it is only when all paths between source and sink are severed, separating the source and sink into two disjoint regions, that the color degrees of freedom of the flux tube produce entanglement entropy.
This topological behavior is manifest in the staggered two-slab geometry with $\Delta y>0$, where the \FTEE of region $V$ at $x\sim0$ generally matches the sum of the \FTEE of its two one-slab components plus a constant vibrational contribution.
When these two subregions are brought into contact ($\Delta y =0$), however, we no longer observe this relation, seeing instead a topological deviation localized around the flux tube.
To see this clearly, we introduce $\tilde{S}_{\text{diff}}$, defined as
\begin{equation}
    \label{eq:sdiff}
    \tilde{S}_{\text{diff}}(x)\equiv \tilde{S}_{A\cup B}(x)-\left(\tilde{S}_{A}(x)+\tilde{S}_{B}(x)\right)\,,
\end{equation}
where $\tilde{S}_{A\cup B}(x)$ is the \FTEE of the $\Delta y >0$ geometry in Fig.~\ref{fig:StaggerGeom} (Center), while $\tilde{S}_{A/B}(x)$ is the \FTEE with $V$ equal to only one of the constituent half-slabs.
$\tilde{S}_{\text{diff}}$ represents the degree to which the \FTEE exceeds the sum of its constituent parts.
$\tilde{S}_{\text{diff}}$ can also be expressed as the difference of mutual information in the presence and absence of a flux tube,  
\begin{equation}
    \tilde{S}_{\text{diff}}=I(A,B)-I_{\vert{\QQbar}}(A,B)\,,
\end{equation}
where $I(A,B)$ is the mutual information defined for regions $A$ and $B$.

\begin{figure}[ht!]
\centering
\includegraphics[width=.49\textwidth,valign=c]{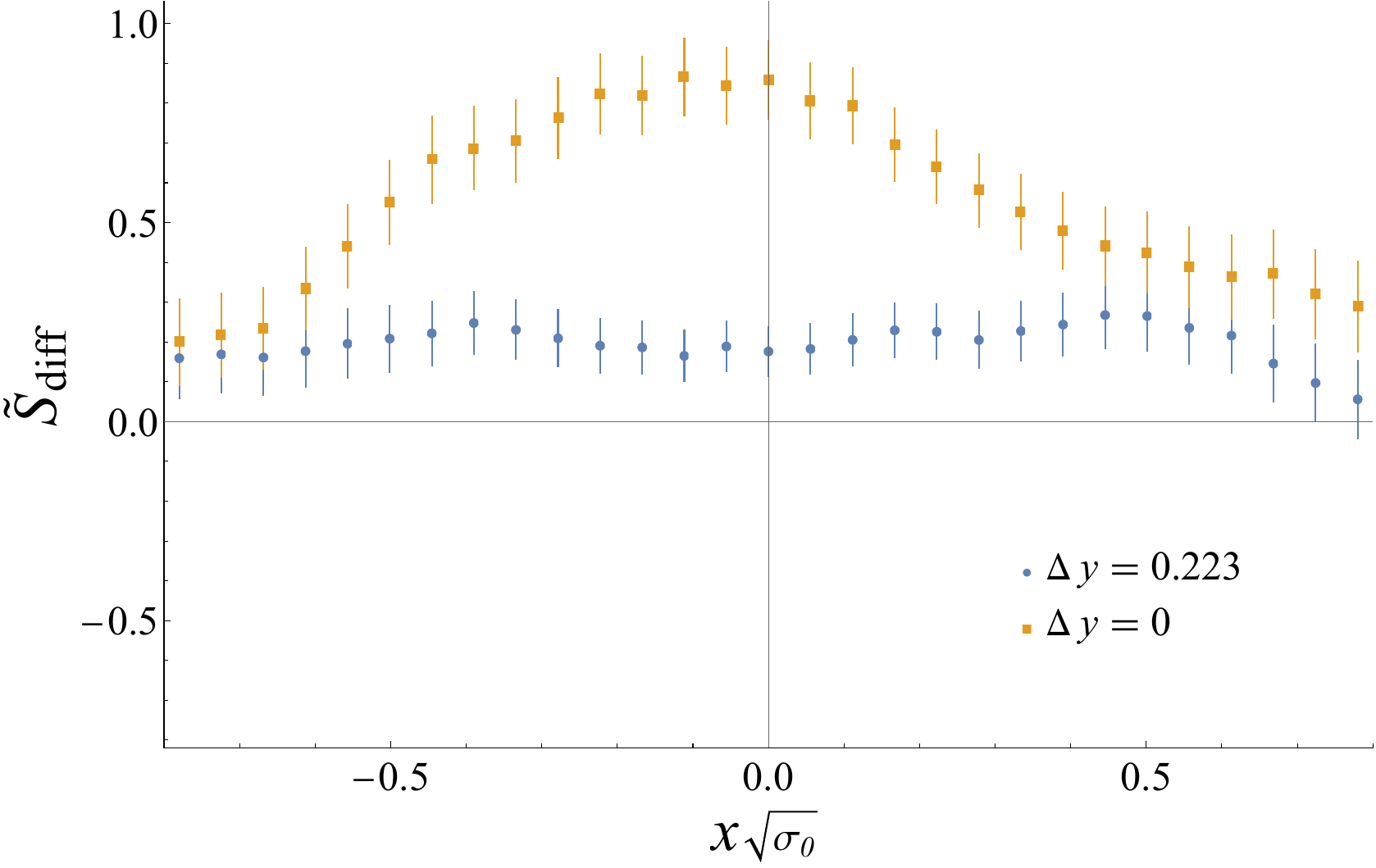}
\caption{\label{fig:sumparts} 
$\tilde{S}_{\text{diff}}$ (defined in Eq.~(\ref{eq:sdiff})) of  the staggered two-slab geometry in $SU(2)$ Yang-Mills as a function of the gap size $\Delta y$. Here $a\sqrt{\sigma_0}=0.056$, the individual slab width is $4a$, and the quark separation $L=20a$.
}
\end{figure}

Only when these two slabs are brought into contact, sealing off all paths from source to sink and changing the topology of $V$ (and by extension, the effective string in $\bar{V}$), do we see $\tilde{S}_{\text{diff}}$ jump from a (roughly) constant value to a form clearly peaked at $x=0$.
The results of this study are shown in Fig.~\ref{fig:sumparts}.\footnote{For disjoint $A$ and $B$, the UV-divergent boundary terms of the entanglement entropy are subtracted in $I(A,B)$, resulting in a finite and universal mutual information.
When $A$ and $B$ share a vertex (as they do for $\Delta y=0$), the mutual information can have a logarithmic divergence due to the UV-cutoff~\cite{Casini:2006hu}.
As seen in Ref.~\cite{Amorosso:2025tgg}, where $\Delta y=0$ and $\Delta y=2a$ configurations are simulated with the coarser lattice spacing of $a\sqrt{\sigma_0}=0.1124$, the discontinuity in \FTEE as region $V$ changes topology is of the same magnitude as seen for the finer lattice spacing studied here ($a\sqrt{\sigma_0}=0.0557$). 
The non-universal boundary terms of the entropy diverge in the continuum limit; therefore, the observed jump, which is insensitive to the UV-cutoff/lattice spacing, must come from the finite portion of the entanglement entropy.}
At the same time, when $\Delta y$ becomes zero, we see \FTEE at $x=0$ jump from roughly 70\% of its maximum to 100\%, the value seen at the $x\to-\infty$ limit of the half-slab geometry.
The staggered two-slab geometry with $\Delta y=0$ shares little in common with the $x\to-\infty$ limit of the half-slab geometry aside from its topology.
The area of $V$ exposed to the flux tube is vastly different, as is its shape.
However, both ensure that the effective string has precisely two boundary crossings (entering and exiting slab $V$), and they have the same \FTEE as a result.

It is important to note that these observations hold regardless of the simplifying assumptions of the naive model we have introduced.
One can mandate all flux tubes have the same entanglement radius $\xi_0$, as the previously stated model does, or some probability distribution for the radius $\xi$, or allow for multiple vibrational modes; whatever model of flux tube entanglement is used, it must be topological in nature to explain these results.

\begin{figure}[ht!]
\centering
\includegraphics[width=.3\textwidth,valign=c]{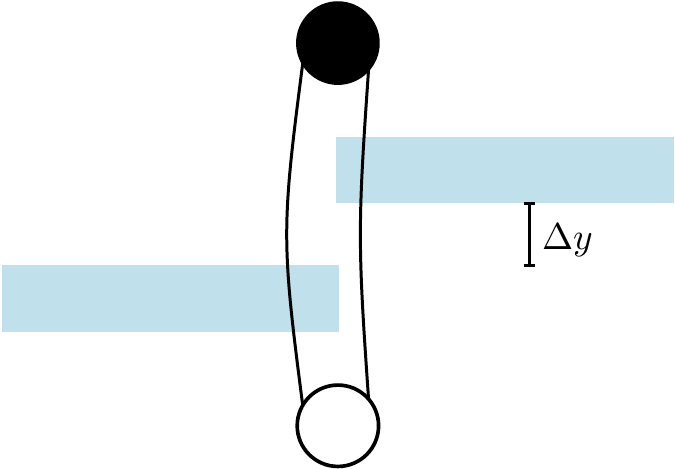}
\hspace{2cm}
\includegraphics[width=.3\textwidth,valign=c]{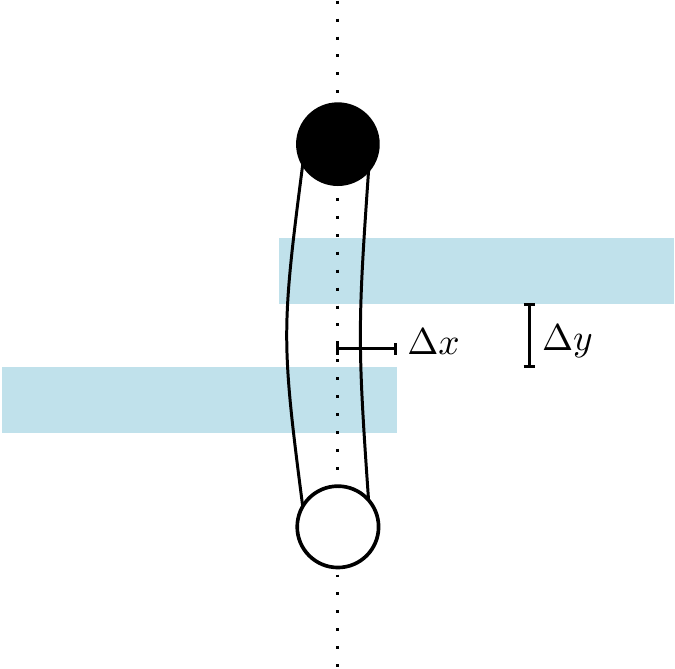}
\caption{\label{fig:deltaXzero} 
Schematics of the staggered two-slab geometry with zero (Left) and positive (Right) $\Delta x$. Note that for the same flux tube deflection, increasing $\Delta x$ can result in a partial boundary crossing ($F=0$) becoming one or more full boundary crossings ($F>0$).}
\end{figure}

To further study this emergent entanglement radius, one can take the $\Delta y >0$ geometry at $\Delta y=0.223$ and slide the slabs relative to each other in the $x$-direction, introducing an overlap of $2\Delta x$, as shown in Fig.~\ref{fig:deltaXzero}.
Firstly, as seen in Ref.~\cite{Amorosso:2025tgg}, when the flux tube is forced to cross two slabs (four total boundary crossings), one 
obtains the predicted $4\ln N_c$ internal entropy contribution.
By studying the overlap $\Delta x$, we can probe the entanglement radius further.
According to our simple model, configurations with four boundary crossings can only occur when the overlap $\Delta x$ is greater than the entanglement radius $\xi_0$.
Conversely, configurations with  partial boundary crossings only occur when $\Delta x<\xi_0$.
This gives us a sensitive probe of the entanglement radius, where $\Delta\tilde{S}\equiv\tilde{S}(x=0)-\tilde{S}(x=\infty)$ is expected to be positive when $\Delta x>\xi_0$ and zero for $\Delta x=\xi_0$.

To get a rough estimate of  model expectations, we once again turn to Eq.~(\ref{eq:roughEst}).
By increasing $\Delta x$, we are decreasing the amount of freedom the flux tube has in order to only exhibit a partial boundary crossing. Instead of the center of the string $\chi$ being bound by the condition $x-\xi_0<\chi<x+\xi_0$ for a partial boundary crossing, it tightens to $x-\xi_0+\Delta x<\chi<x+\xi_0 -\Delta x$. This causes the dip in the profile of \FTEE to gradually vanish as $\Delta x$ increases, eventually reversing and becoming a small bump: When $x+\xi_0-\Delta x<\chi<x-\xi_0 +\Delta x$, the effective string exhibits four full boundary crossings instead of two, contributing $4\ln2$ \FTEEformat.
The difference between $\tilde{S}(x=0)$ and $\tilde{S}(x=\infty)$ is a useful metric, and according to these simplifying assumptions, it should take the form
\begin{equation}
\label{eq:roughEst2}
    \Delta \tilde{S}_{Q\bar{Q}}\equiv\tilde{S}_{Q\bar{Q}}(x=0)-\tilde{S}_{Q\bar{Q}}(x=\infty)= 
 (2\ln 2)\cdot\left[\Phi\left(\frac{\Delta x-\xi_0}{w}\right) -\Phi\left(\frac{-\Delta x+\xi_0}{w}\right)\right]\,.
\end{equation}
\begin{figure}[ht!]
\centering
\includegraphics[width=.49\textwidth,valign=c]{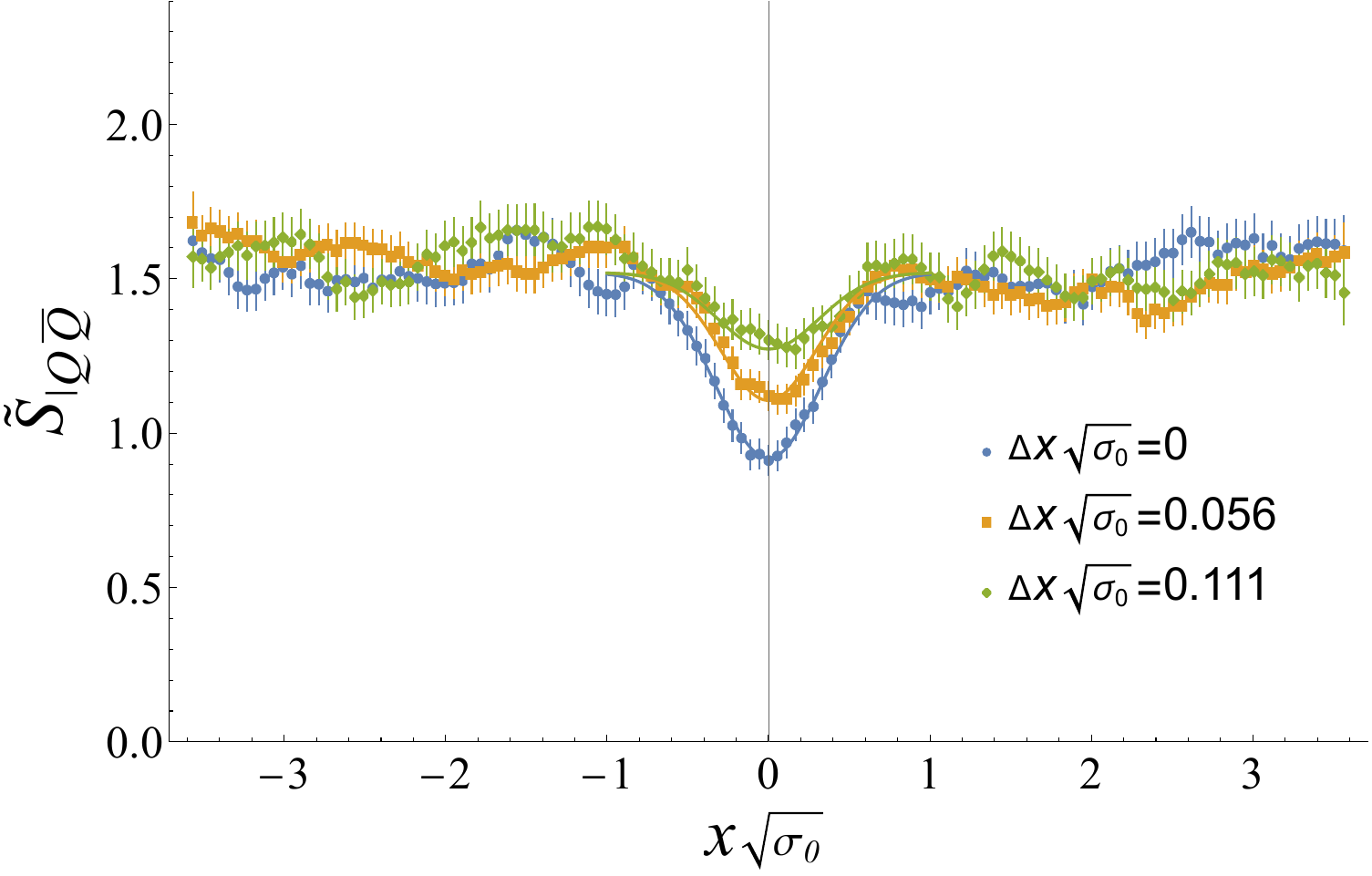}
\includegraphics[width=.49\textwidth,valign=c]{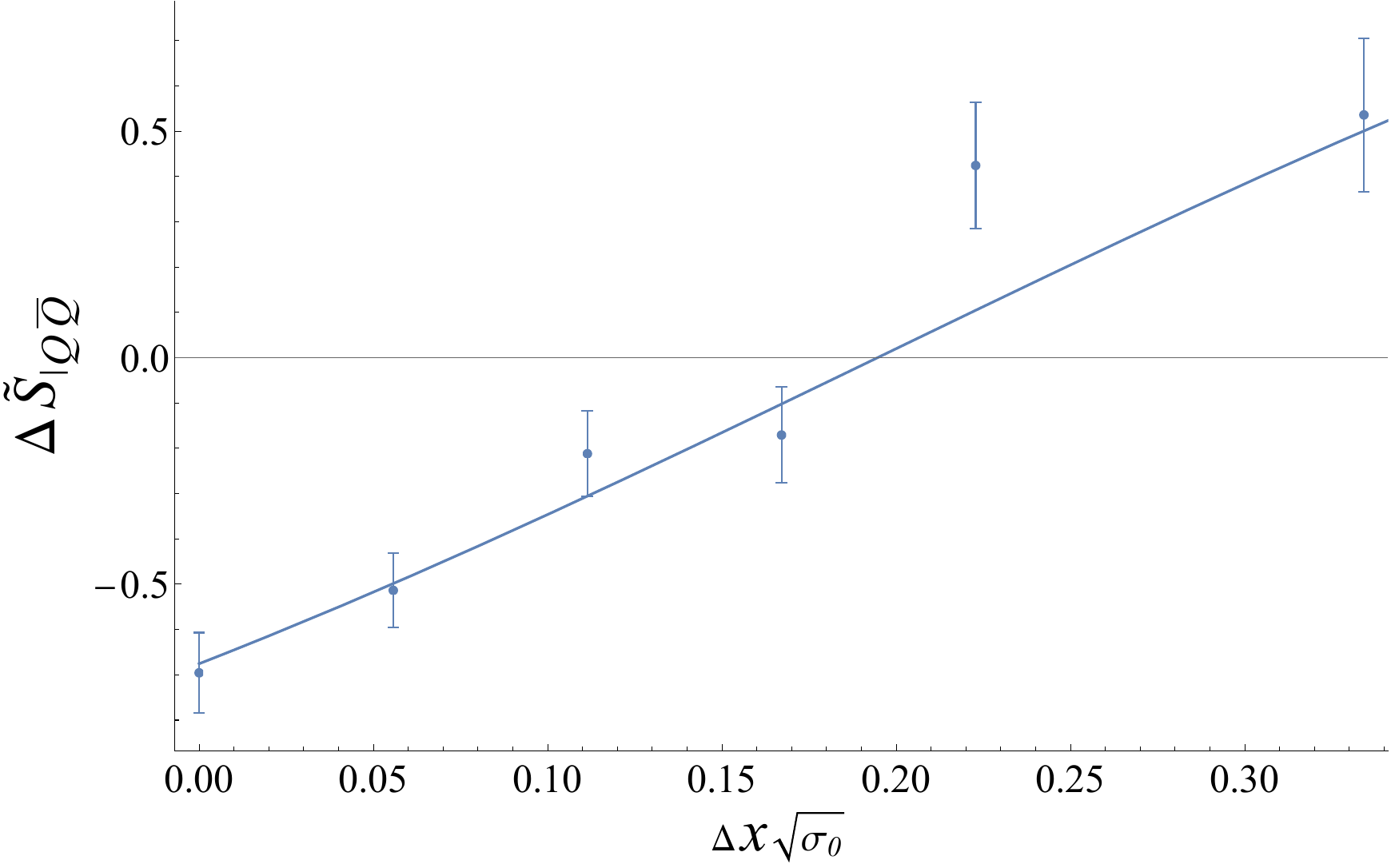}
\caption{\label{fig:dubtest2stagger2} 
(Left) Comparison of the \FTEE of the staggered two-slab geometries with different $\Delta x$ overlaps in $SU(2)$ Yang-Mills. Note that the dip at $x=0$ becomes less pronounced as $\Delta x$ increases.
(Right) The difference between \FTEE at $x\to\pm\infty$ and $x=0$. In both figures, $a\sqrt{\sigma_0}=0.056$, the individual slab width is $4a$, $\Delta y = 4a$, and the quark separation $L=20a$. Solid lines represent the fits to Eq.~(\ref{eq:roughEst}) (left) and Eq.~(\ref{eq:roughEst2}) (right), with a fitted constant added to Eq.~(\ref{eq:roughEst}) to account for vibrational contributions.}
\end{figure}

The prediction of the model again matches what we see on the lattice, with the $x$-value at which $\Delta \tilde{S}=0$, $\xi_{0,\text{flat}}=0.195(29)$, matching the offset and fitted values within uncertainty.
Once again, we see three geometries of $V$, the overlapping slab geometry with $\Delta x=\xi_0$, the staggered two-slab geometry with $\Delta y=0$, and the half-slab geometry at $x\to-\infty$, which only have one feature in common: They always split the portion of the effective string in $\bar{V}$ into two separate components.
These three geometries share the same internal component of \FTEE due to this fact, providing further evidence that the entanglement entropy of the flux tube behaves in the conjectured topological fashion.

\begin{figure}[ht!]
\centering
\includegraphics[width=.49\textwidth,valign=c]{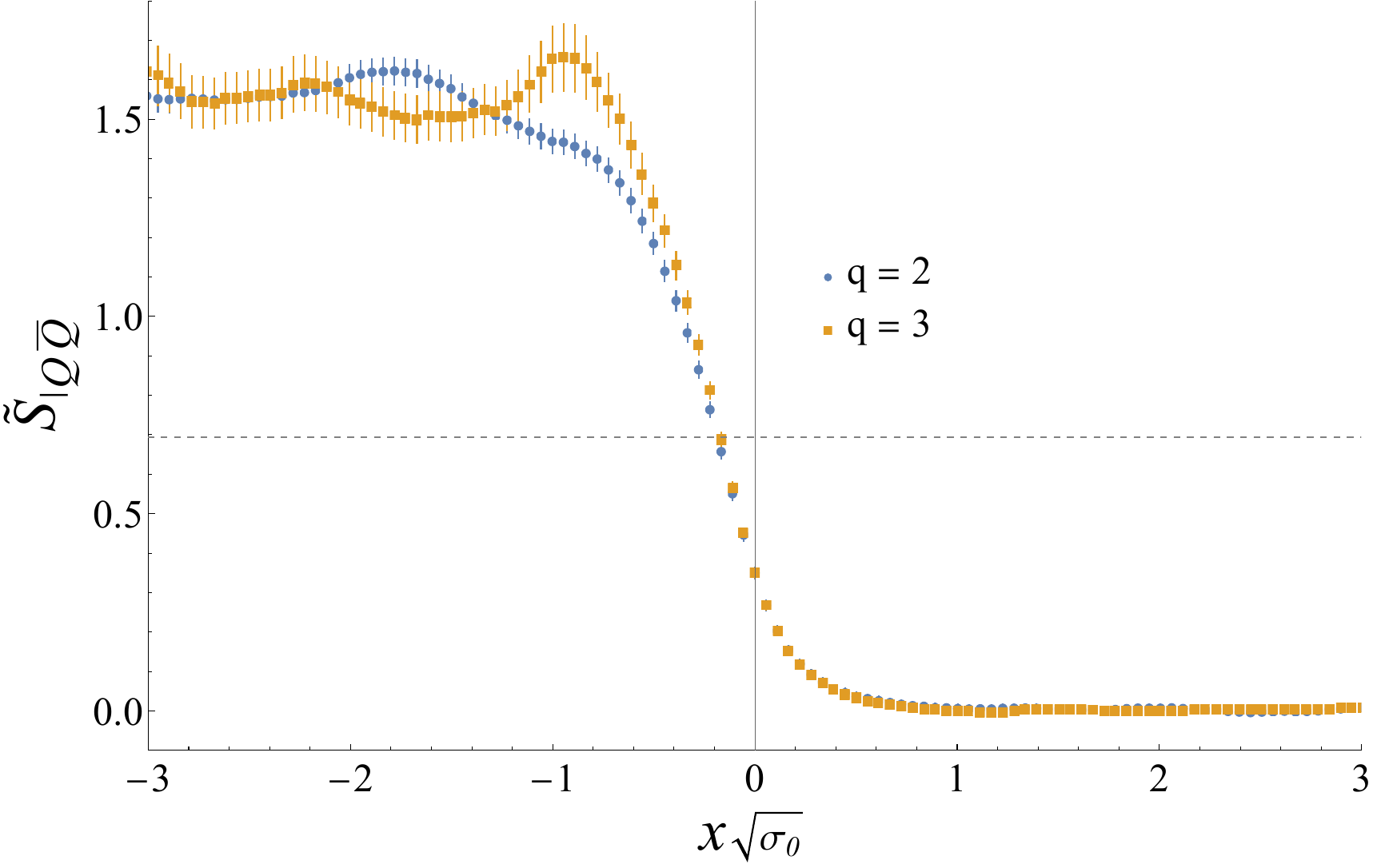}
\includegraphics[width=.49\textwidth,valign=c]{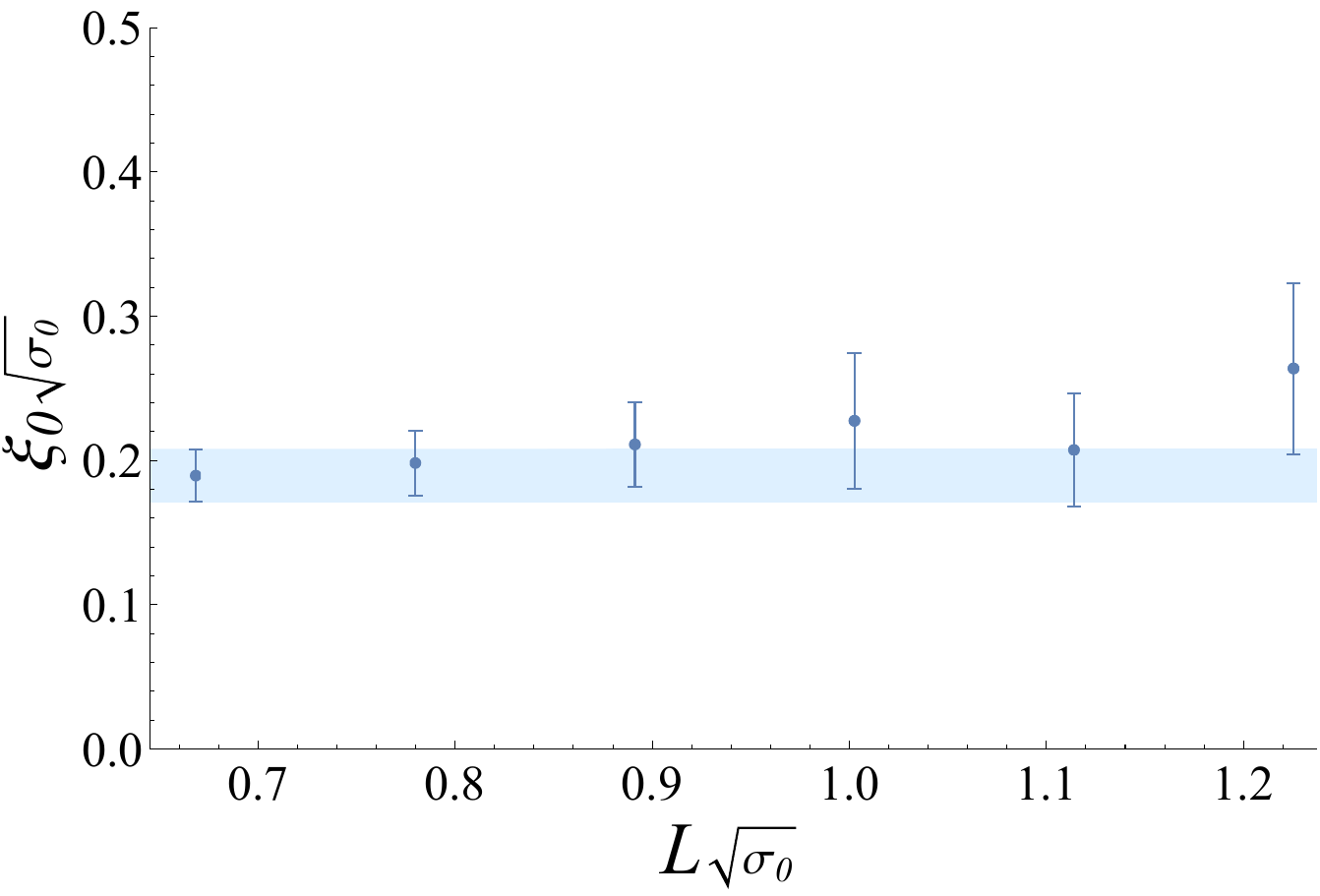}
\caption{\label{fig:effectiveIntrinsic} 
(Left) \FTEE of the half-slab geometry in $SU(2)$ Yang-Mills for an $L\sqrt{\sigma_0}=0.67$ flux tube with $q=2,3$.
The dashed line denotes $\tilde{S}=\ln{2}$, corresponding to $F\approx1$. Note both plots reach $\tilde{S}=\ln{2}$ at roughly the same point. 
(Right) The entanglement radius $\xi_0$, measured by the offset method as a function of the inter-quark separation $L$.
The bands are a visual aid representing the error bars at $L\sqrt{\sigma_0}=0.67$.
For both figures $a\sqrt{\sigma_0}=0.056$.
}
\end{figure}

Before studying this entanglement radius further, we should confirm that it behaves as a distinct physical property  of the flux tube's internal structure.
Physical scales associated with the flux tube's internal structure such as the intrinsic width~\cite{Caselle:2012rp,Verzichelli:2025cqc,intrinsic:inprep} and the London penetration depth should not depend on the length of the flux tube; we therefore expect a similar result for the entanglement radius.
Further, the entanglement radius characterizes the quantum state of the fields, and its determination employing the R\'{e}nyi entropy of any order $q$ should yield the same value, as well as employing the von Neumann entropy (although we do not have means to test the latter).
We showed explicitly in Ref.~\cite{Amorosso:2024glf} that the internal component of the entanglement entropy in (1+1)D Yang-Mills theory does not depend on $q$; it only depends on the number of boundary crossings and the dimension of the representation of the quark sources.
As the entanglement radius is derived from the number of full boundary crossings and their internal entropy contributions, this tells us that its  value should not depend on $q$ either.
Comparisons of $\xi_0$ values determined with $q=2$ and $q=3$ and varying $L$ are presented in Fig.~\ref{fig:effectiveIntrinsic}, which show that they do not have an effect on the entanglement radius.\footnote{
As mentioned earlier, one may argue against the topological model by assuming an unphysical energy cost to transporting flux across a boundary in a replica geometry, artificially suppressing thin string configurations that cross boundaries. This is ruled out by the lack of $q$ and $L$-dependence observed, as the suppression would cause higher $q$ and $L$ to be associated with larger offsets of \FTEEformat.
}

We have now presented compelling evidence, through several measures,  that the flux tube has finite entangling width that must be fully cross-cut to contribute to the internal entanglement entropy.
This finding has wide ramifications and allows us to reinterpret our previous results.
For the remainder of the paper, we will focus  on the half-slab geometry (Fig.~\ref{fig:refinedhalfslab}) to examine how the topological model of flux tube entanglement provides insight into the flux tube's internal structure.

\begin{figure}[ht!]
\centering
\includegraphics[height=.3\textwidth,valign=c]{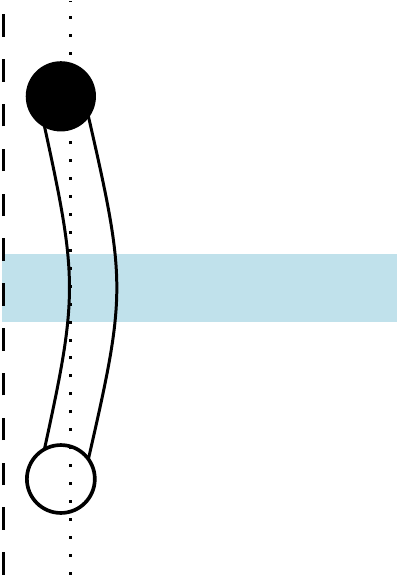}
\hspace{2.5cm}
\includegraphics[height=.3\textwidth,valign=c]{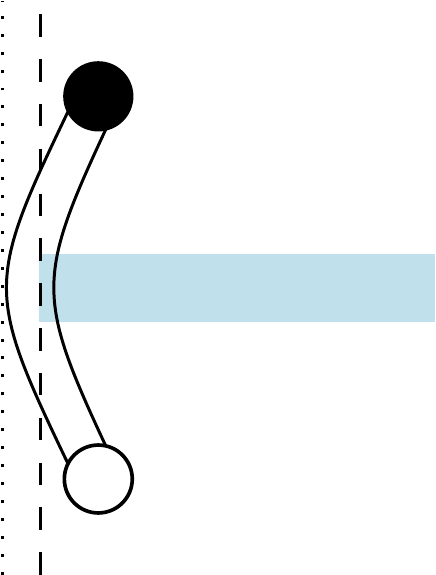}
\hspace{2cm}
\includegraphics[height=.3\textwidth,valign=c]{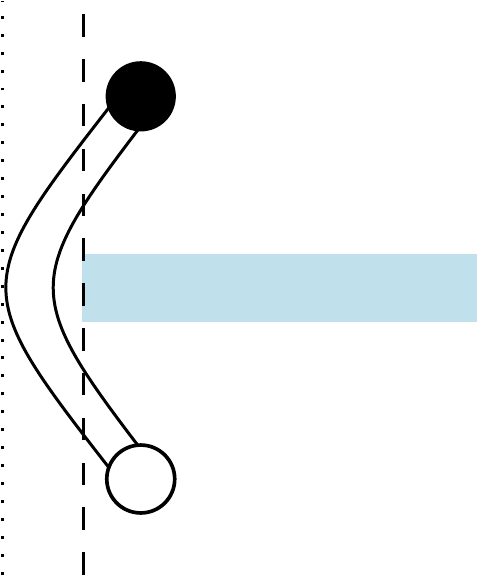}
\caption{\label{fig:BoCrWide} 
Schematic of three flux tubes with differing deflections such that they fully intersect region $V$ (Left), partially intersect region $V$ (Center), and do not intersect region $V$ at all (Right).
Region $V$ is shaded blue.
For a sufficiently thin slab, the condition of a ``full boundary crossing'' is equivalent to $\chi_L$ (dotted line) being greater than $x$ (dashed line).
}
\end{figure}

In the half-slab geometry, the flux tube must deflect sufficiently to the right ($\chi>0$) to fully intersect region $V$ and produce internal entropy.
As only full boundary crossings contribute to \FTEEformat, the only relevant parameter when determining whether a flux tube configuration has nontrivial internal entropy is then the location $\chi_L=\chi-\xi_0$ of the left wall of the flux tube in Fig.~\ref{fig:refinedhalfslab}: If $\chi_L>x$, both walls will be inside region $V$, the entanglement radius will be fully severed, and the configuration will have \FTEE of $2\ln N_c$ as a result.
If $\chi_L<x$, the left wall will be outside region $V$, and the flux tube will either exhibit a partial boundary crossing ($\chi_R=\chi+\xi_0>x$) or miss $V$ entirely ($\chi_R<x$).
In either case, the \FTEE contribution will be zero, and the location of the rightmost wall $\chi_R$ is irrelevant.
The full, partial, and no boundary crossing scenarios are depicted in Fig.~\ref{fig:BoCrWide}.
Given this, \FTEE in the half-slab geometry is essentially the cumulative distribution function of $\chi_L$, rescaled by a factor of $2\ln N_c$.
Our simple model of the vibrating effective string with Gaussian-distributed deflection and entanglement radius $\xi_0$ would predict an error function shape of \FTEEformat, symmetric about $x=-\xi_0$ instead of $x=0$.
This is of course what we saw in Ref.~\cite{Amorosso:2024leg}, motivating the topological interpretation of flux tube entanglement.

\begin{figure}[ht!]
\centering
\includegraphics[width=.49\textwidth]{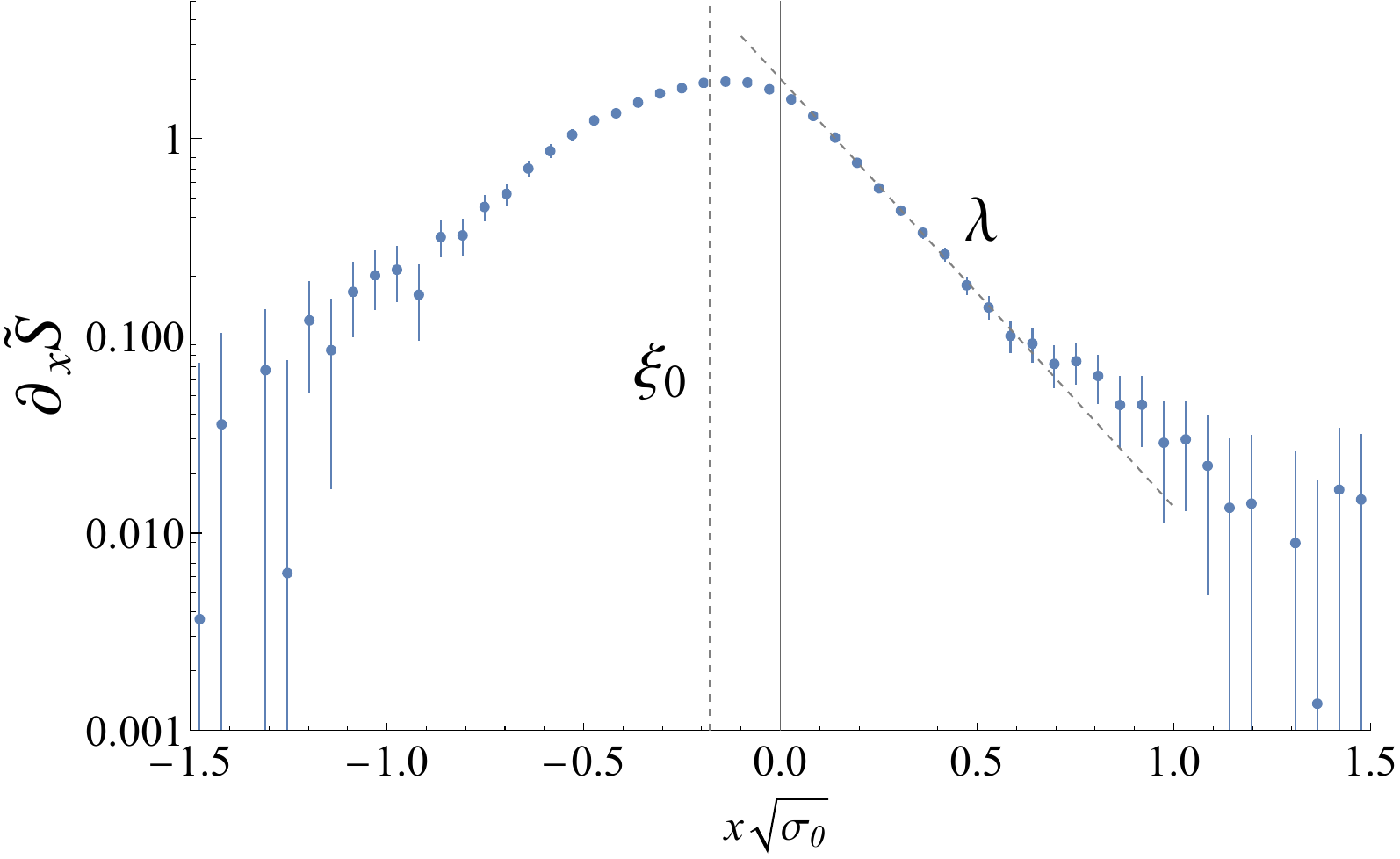}
\caption{\label{fig:lambdaxi}
Log-plot of the derivative of \FTEE of the half-slab geometry in $SU(2)$ Yang-Mills with $a\sqrt{\sigma_0}=0.056$ and inter-quark separation $L=12a$. We observe exponential falloff and an offset, governed by parameters $\lambda$ and $\xi_0$, respectively. The profile is centered about $x=-\xi_0$ and decays with rate $\propto e^{-x/\lambda}$.
}
\end{figure}

We can further refine the model by studying the derivative of \FTEEformat. 
The statistical noise in \FTEE values at nearby points $x$ computed on a lattice is strongly correlated; as a result, the difference between the \FTEE values of close points has better statistical precision compared to the \FTEE values themselves.
We can then use the derivative of \FTEE to study $P(\chi_L)$, the probability distribution function of $\chi_L$, with high precision.
Doing so reveals that $P(\chi_L)$ is actually not Gaussian, as would be expected for a fixed entanglement radius.
It instead takes a form that is Gaussian around its peak located approximately at $\chi_L=-\xi_0$, but decays exponentially $\propto e^{-\vert\chi_L\vert/\lambda}$ in its ``tails'' $|\chi_L|\gg\xi_0$.
The plot of $\partial_x \tilde{S}_{\vert\QQbar}$, the $x$-derivative of \FTEE in the half-slab geometry, is shown in Fig.~\ref{fig:lambdaxi}. 
This exponential decay is indicative of a richer model of the flux tube's internal structure than the naive model of a fixed entanglement radius $\xi_0$ discussed earlier.

To understand the implications of this exponential decay, it is instructive to first briefly consider a simpler well-understood system with similar dynamics: the electric field of (2+1)D compact $U(1)$ gauge theory, which has the same dynamics as QED$_3$ at low energy scales below the electron mass.
In compact $U(1)$ gauge theory, one can study the electric field in the presence of static quark sources as a function of the transverse displacement $y$ from the inter-quark axis. One arrives at a similar distribution to Fig.~\ref{fig:lambdaxi}, albeit centered about $y=0$.
This form, a distribution with a Gaussian peak and exponential tails, is understood in this compact $U(1)$ case to be a consequence of both the effective string's deflection and the intrinsic width of the flux tube~\cite{Aharony:2024ctf}.
While the effective string theory of the flux tube in QED$_3$ exhibits Gaussian deflection, QED$_3$ can also be expressed in the Villain form for scales below the electron mass, from which a classical soliton solution of the dual photon can be derived~\cite{Aharony:2024ctf}
\begin{equation}
   \phi_{cl}=\frac{2e}{\pi}\text{arctan}(e^{-m(x-x_0)}) \,,
\end{equation}
where $\phi_{cl}$ represents a flux tube centered at $x_0$, $e$ is the gauge coupling constant, and $m$ is the dual photon mass which determines the width of the soliton.
As is done in Ref.~\cite{Aharony:2024ctf}, one can fix the string at $x=x_0$ and expand $\phi$ about $\phi_{cl}$, arriving at the solution
\begin{equation}
\label{eq:NGBsoliton}
    \phi_0=\sqrt{\frac{m}{2}}\text{sech}(mx)\,
\end{equation}
describing the Nambu-Goldstone boson bound to the string.
This function, proportional to the derivative of the soliton solution $\phi_{cl}$ (and therefore its electric field as well), decays at the same rate as the transverse electric field on the lattice.
In fact, lattice calculations of the transverse electric field can be accurately parameterized by a convolution of two distributions\footnote{While the lattice measurements in Ref.~\cite{Caselle:2016mqu} can be parameterized by a convolution of the two distributions mentioned, these measurements are outside the stated region of validity of the effective theory presented in Ref.~\cite{Aharony:2024ctf}, which requires both $m^2/\sigma$ and $\sigma a^2$ to be small; $m^2/\sigma$ is of order unity in Ref.~\cite{Caselle:2016mqu}.}: the probability distribution function corresponding to Gaussian deflection of the effective string and the expression $\phi_0$ as seen in Eq.~(\ref{eq:NGBsoliton})~\cite{Aharony:2024ctf,Caselle:2016mqu}.
The dual photon mass $m$ can be extracted from the exponential decay of the electric field at large transverse displacement where the Gaussian part of the convolution has already decayed.
This has been done for $SU(2)$ as well, extracting a scale dubbed the intrinsic width~\cite{Verzichelli:2025cqc,Caselle:2012rp,intrinsic:inprep}.

In QED$_3$, the Gaussian profile can be regarded as the wave function of the string, while the exponentially-decaying soliton solution can be thought of as the intrinsic profile of the flux tube dictating its intrinsic width.
The intrinsic width can be conceptualized similarly to the Compton wavelength of an electron~\cite{Aharony:2024ctf}.
In the case of an electron, its observed density will depend on both its wave function (which in turn depends on the potential it is subject to) as well as its Compton wavelength.
Analogously, the flux tube's electric field depends both on its effective string description (which in turn depends on the source/sink separation) as well as its intrinsic width.

We apply similar logic to the Yang-Mills color flux tube, and study its structure using the $x$-derivative of \FTEEformat.
The probability distribution of the location of the flux tube's leftmost wall  $P(\chi_L)$ should analogously be a convolution of two distributions: the Gaussian distribution of the effective string deflection and the distribution $P_0(\chi_L)$ of the leftmost wall assuming the string deflection is zero.
Just as the electric field in QED$_3$ depends on the source/sink separation as well as the dual photon mass, it is natural that \FTEE depends both on the static quark separation as well as the intrinsic width of the flux tube.

\begin{table}[ht!]
\centering
\begin{tabular}{c|c|c|c|c}
\hline\hline
\textbf{Group} & \textbf{$\xi_0\sqrt{\sigma_0}$} & \textbf{$\lambda\sqrt{\sigma_0}$} & $\sqrt{\sigma_0}/m_{0^{++}}$\\
\hline
$SU(2)$ & 0.185(6) & 0.223(15) & 0.212(2)\\
$SU(3)$ & 0.269(8) & 0.199(45) & 0.231(2)\\
$SU(4)$ & 0.321(8) & 0.218(31) & 0.236(3)\\
$SU(5)$ & 0.393(9) & 0.191(29) & 0.239(3)\\
\hline\hline
\end{tabular}
\caption{Comparison of $\xi_0$ (as extracted through the ``offset method,'' which  calculates the $x$-value for which \FTEE equals $\ln N_c$) and $\lambda$ for various $N_c$ at lattice spacing $a\sqrt{\sigma_0}=0.06$.
The inverse glueball mass from Ref.~\cite{Teper:1998te} is also displayed for reference.}
\label{table:params}
\end{table}

Information about $P_0(\chi_L)$ can be extracted as follows:
since the Gaussian deflection of the string is symmetric about $x=0$, 
while $\partial_x\tilde{S}_{\vert\QQbar}$ has exponential decay in its tails and is not symmetric about $x=0$, both features must arise from $P_0(\chi_L)$.
Thus, we can extract two parameters of $P_0(\chi_L)$ by studying $\partial_x\tilde{S}$: the familiar offset $-\xi_0$, as well as the decay rate $\lambda$.
These parameters, illustrated in Fig.~\ref{fig:lambdaxi}, are summarized in Table~\ref{table:params} for $SU(N_c)$ gauge theories with $N_c=2\ldots5$. 
These findings imply a significantly more complex model of the flux tube's internal structure than the simple model of fixed entanglement radius discussed previously.
Specifically, there exists a probability distribution $P_0(\chi_L)$, a conditional distribution $P_0(\chi_R\vert \chi_L)$, and the implied distribution of entanglement radii that can be derived from these two.
The latter two distributions can be investigated through lattice studies where the entangling region is set to length $L_{V}\sim\xi_0$.
Doing so reveals that the entanglement radius is not fixed, but follows a distribution with mean $\xi_0$.\footnote{We will provide these results with finer detail extracted from higher statistics lattice data in an upcoming paper.}
However, we leave the task of deriving a fully robust effective string theory for future work, and opt to solely study parameters $\xi_0$ and $\lambda$ of $P_0(\chi_L)$ in this work.
Despite these complications, note that $\xi_0$ carries the same meaning as discussed previously.
Indeed, the average entanglement radius $\frac{1}{2}\langle \chi_R-\chi_L\rangle=\frac{1}{2}\langle \chi_R \rangle - \frac{1}{2}\langle \chi_L\rangle = -\langle \chi_L\rangle=\xi_0$, where the expectation value is taken according to the distribution $P(\chi_{L/R})$ (or equivalently, $P_0(\chi_{L/R})$).

\begin{figure}[ht!]
\centering
\includegraphics[width=.49\textwidth]{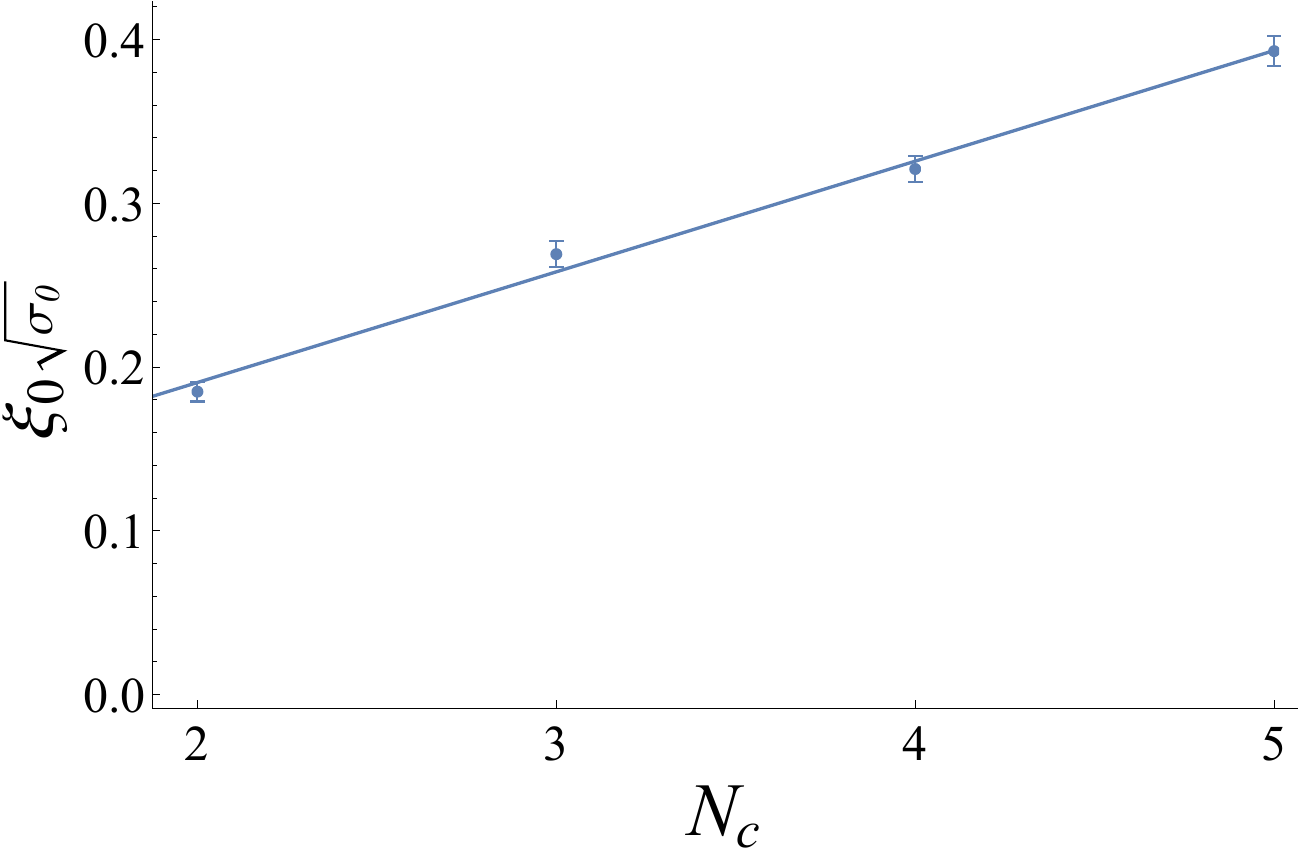}
\includegraphics[width=.49\textwidth]{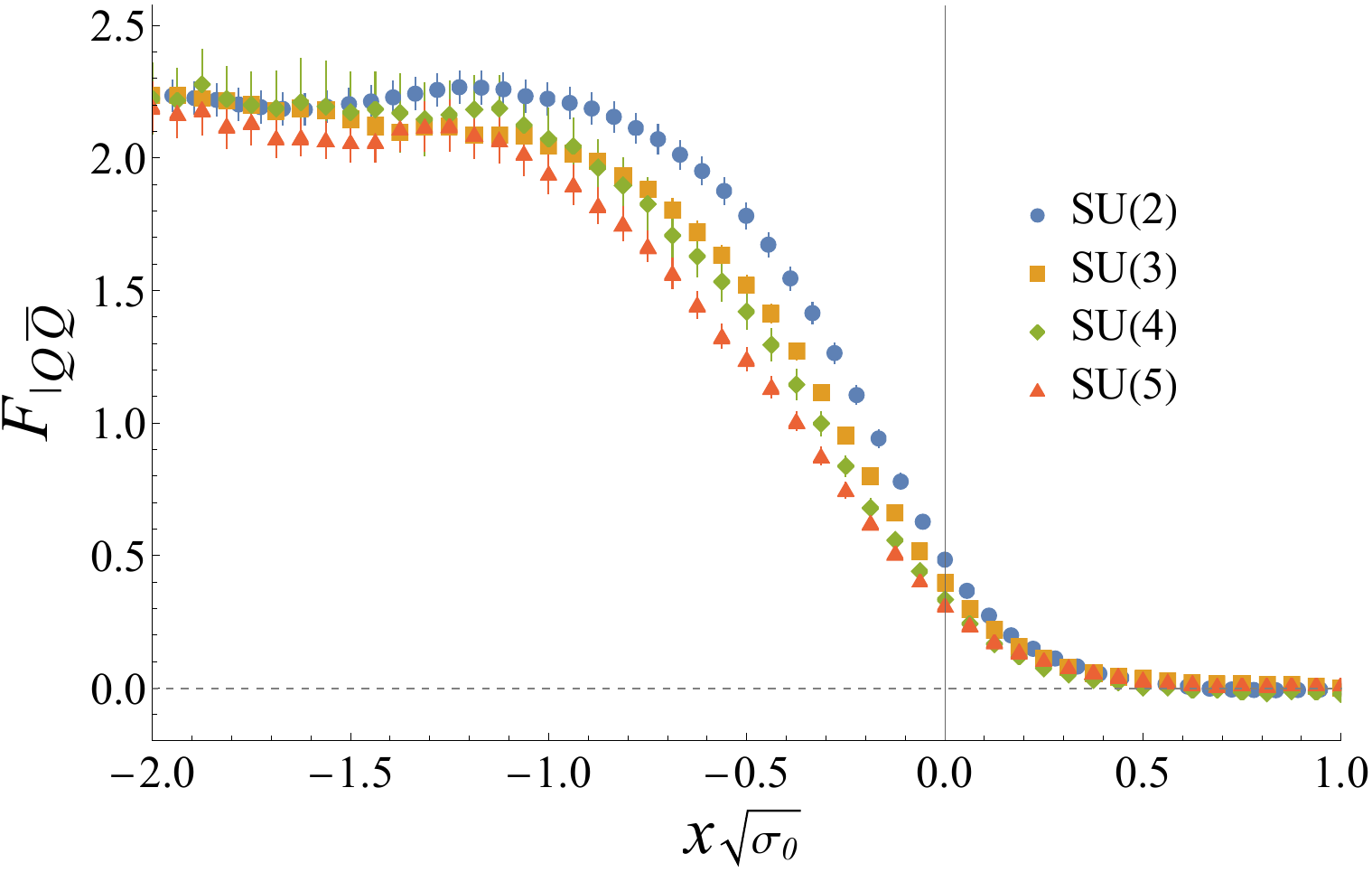}
\caption{\label{fig:ncDependence} 
(Left) Entanglement radius (extracted using offset method) as a function of $N_c$. For $N_c=2,3$, the entanglement radius at this lattice spacing is approximately at its continuum value; however, for $N_c=4,5$, significant discretization effects may remain (see Figs.~\ref{fig:iwscaling} and \ref{fig:iwscalinglargeN}).
(Right) \FTEE of the half-slab geometry normalized by $\ln N_c$, where $F\equiv \tilde{S}/\ln(N_c)$. 
The profiles are similar but shifted by the entanglement radius.
Lattice data obtained for both figures with an $L=12a$ quark separation and $a\sqrt{\sigma_0}=0.06$.
}
\end{figure}

\begin{figure}[ht!]
\centering
\includegraphics[width=.49\textwidth]{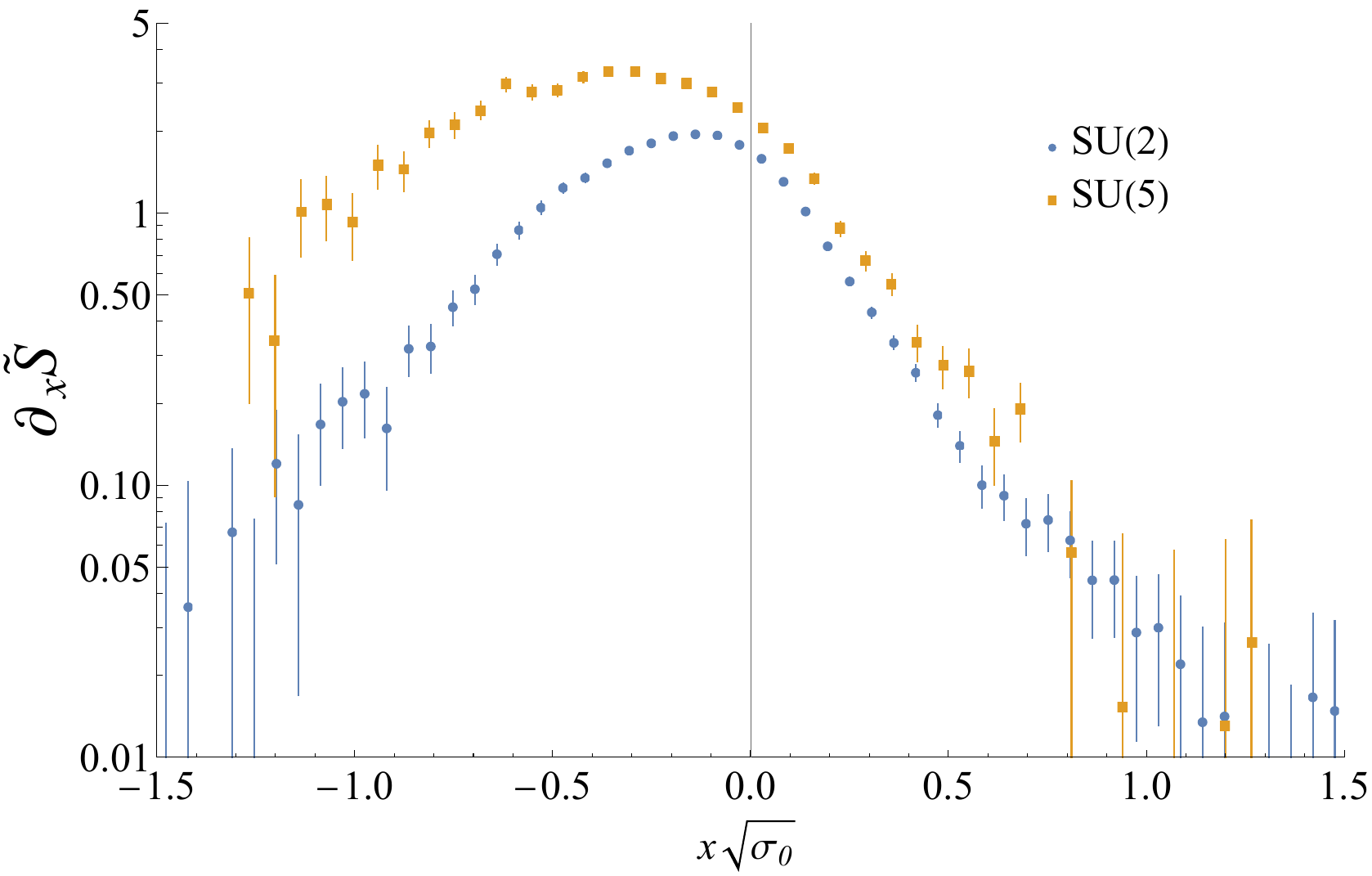}
\caption{\label{fig:su2su5}
The derivative of \FTEE as a function of $x$ with an $L=12a$ quark separation and $a\sqrt{\sigma_0}=0.06$.
We see that the right tail behavior, governed by scale $\lambda$, is similar for gauge groups $SU(2)$ and $SU(5)$.
The left tail behavior is contaminated by unknown vibrational contributions yet decays with similar scale.
Note that the profiles for different gauge groups are largely similar but rescaled by $\ln N_c$ and shifted by the entanglement radius.
This behavior is consistent for $N_c=3,4$ as well, but they have been omitted for clarity.
}
\end{figure}

Comparing $\partial_x\tilde{S}$ for gauge groups $SU(2)$ through $SU(5)$, one can see $\xi_0\sqrt{\sigma_0}$ grows linearly with the number of colors while $\lambda\sqrt{\sigma_0}$ is roughly constant as a function of $N_c$, with the latter mirroring the behavior of the inverse mass of the lightest glueball.
A scaling study for $\xi_0$ has been performed for gauge groups $SU(2)$ through $SU(5)$, and can be seen in Figs. \ref{fig:iwscaling} and \ref{fig:iwscalinglargeN}.
Our calculated values of $\lambda$ agree with the inverse mass of the lightest glueball for $SU(2)$ through $SU(4)$, while $SU(5)$ produces a value of $\lambda$ that is slightly more than one standard deviation away from $1/m_0$.
This behavior is reminiscent of the transverse electric field in QED$_3$, where the rate of exponential decay of the electric field is equal to the inverse dual photon mass.
The approximately linear dependence of $\xi_0$ on $N_c$ is plotted in Fig.~\ref{fig:ncDependence}, and $\partial_x\tilde{S}$ is compared between $SU(2)$ and $SU(5)$ in Fig.~\ref{fig:su2su5} as functions of $x\sqrt{\sigma_0}$.
One can see that the overall profile of $\partial_x\tilde{S}$ remains the same as the $N_c$ increases, while the offset appears to increase linearly.
The origin of this behavior is mysterious; while the decay rate $\lambda$ seems to be associated with the inverse mass of the lightest glueball and the intrinsic width of the flux tube in $SU(2)$ Yang-Mills~\cite{Verzichelli:2025cqc,intrinsic:inprep}, the seemingly linear scaling of $\xi_0$ with $N_c$ has no immediate explanation.\footnote{As a result, it is unclear how measurements of $\xi_0$ through \FTEE can be related to measurements of the transverse electric field without knowing $P_0(\chi_L,\chi_R)$ at high precision.}

Understanding of the $N_c$ systematics of this novel entanglement radius can provide a deeper understanding of the relation of 
entanglement to confinement in (2+1)D Yang-Mills theory. We will return to this problem in future work where we will also explore this scale in the context of (3+1)-D Yang-Mills theory. 

\section{Summary and Conclusions}
\label{sec:Conclusions}
\FloatBarrier
We demonstrated in this work that flux tube entanglement entropy, introduced previously in Ref.~\cite{Amorosso:2024leg}, is a powerful tomographic tool for studying the color flux tube.
In particular, \FTEE allows us to study the systematics of the flux tube's color degrees of freedom and their contribution to the entanglement entropy.
Our results show that flux tubes possess a rich internal structure that cannot be captured by a 1-D vibrating effective string. 
In $SU(2)$ (2+1)D lattice Yang-Mills theory, we explored the \FTEE of regions $V$ consisting of two disconnected subregions.
We found that \FTEE depends on the topology of region $V$, changing sharply when the two subregions are brought into contact, fully severing flux tubes that pass between them.
This discontinuity defines a novel physical scale in (2+1)D Yang-Mills theory,  the entanglement radius $\xi_0$. It is a topological object, defined to be the minimal width of the entangling region required to fully sever the color flux tube. 
We verified through multiple single-- and multi--slab geometries that $\xi_0$ is independent of its measurement method, the inter-quark separation $L$, and the R\'{e}nyi index $q$.

We also performed lattice simulations of \FTEE for gauge groups $SU(2)$ through $SU(5)$ with the half-slab geometry as in Ref.~\cite{Amorosso:2024leg}.
By varying the slab's transverse position relative to the $\QQbar$ axis, we extracted two scales from the derivative of \FTEEformat: the entanglement radius $\xi_0$, and a decay rate $\lambda$ governing the exponential falloff of the tails. 
We find $\lambda$ roughly constant across gauge groups, consistent with the inverse glueball mass and the intrinsic width~\cite{Verzichelli:2025cqc,intrinsic:inprep}.
Surprisingly, $\xi_0$ grows approximately linearly with $N_c$: We currently have no explanation for this behavior. 

Our work also raises two related open questions.
First, why does the \FTEE of a fully severed flux tube scale as $F\ln N_c$ in the continuum limit, where $F$ is the number of full boundary crossings?
As discussed in the introduction, the topological behavior of \FTEE suggests that it arises from distillable bulk entanglement rather than boundary edge modes.\footnote{Entanglement due to edge-modes can behave topologically in some contexts; in topological $\mathbb{Z}_2$, nonlocal constraints due to flux conservation cause the entanglement to depend on the number of boundaries~\cite{Casini:2013rba}.
We expect similarly for the staggered two-slab geometry explored in Section~\ref{sec:EntanglementRadius}: the change from a single connected boundary at $\Delta y=0$ to a two-component boundary at $\Delta y>0$ will result in a change $\propto\ln(2)$ of the vacuum entanglement entropy.
However, we expect these constraints to contribute equally in vacuum and in presence of a flux tube, leaving \FTEE unaffected by edge mode terms.}
It is unclear why \FTEE remains roughly constant as we move from coarse to fine lattice spacing when the source of the excess entanglement changes.
Second, why does \FTEE behave topologically at all?
We can see similar behavior in a toy model: in (1+1)D spatially periodic Yang-Mills at finite temperature, region $V$ needs to sever both paths from source to sink to have non-trivial \FTEE at weak bare coupling (high temperature).
However, the \FTEE of this model arises from non-distillable entanglement, making its generalization to higher dimensions unclear.
A theoretical framework explaining both the $F\ln N_c$ scaling and the topological behavior of \FTEE remains an important open problem that we will address in future works.\footnote{The observed behavior of \FTEE bears resemblance to the topological entanglement entropy introduced in Refs.~\cite{Kitaev:2005dm,Levin:2006zz}. Whether a direct connection exists remains to be understood.}

Beyond addressing these open questions, the study of  \FTEE can be extended in several directions. One is with adjoint quark sources, identifying the scale at which gluons screen all but $N$-ality.
This will provide insight into the entangling dynamics of a system that undergoes string breaking.\footnote{This can also be accomplished by adding dynamical sea quarks. Studies of the Schwinger model have shown that as inter-quark separation grows, the reduced state of a subsystem reproduces thermal features associated with hadronization~\cite{Berges:2017zws,Grieninger:2025rdi}.
The quantum complexity of string breaking in the Schwinger model has been studied in depth in Ref.~\cite{Grieninger:2026bdq}.
It would be interesting to observe similar dynamics in $SU(N)$ Yang-Mills.}
This would also enable us to probe the landscape of Yang-Mills theories further.
The internal entropy scales as $F\ln N_c$ for fundamental sources in both (1+1)D and (2+1)D $SU(N)$ Yang-Mills, but it is unknown to what extent \FTEE results in (1+1)D translate to higher dimensions.
Adjoint quark sources would allow us to probe whether the general relation $\tilde{S}=F\ln dimR$ observed in (1+1)D, where $dimR$ is the dimension of the quark source's representation, is followed in higher dimensions as well.
Similarly, the scaling of the entanglement radius with $dimR$ of the quark sources could give us further insight into the topological nature of \FTEEformat.

Another important direction is studying the entangling properties of static baryon junctions.
Studying \FTEE with static quark-antiquark sources has allowed us to identify the vibrational width of the emergent color flux tube as well as its entanglement radius and decay rate.
We expect \FTEE to similarly reveal the vibrational width, entanglement radius, and decay rate of the emergent baryon junction.
By studying the vibrational width and intrinsic properties of the baryon junction, we can corroborate recent studies of the mass and dynamics of the junction in effective string theory~\cite{Komargodski:2024swh} and on a lattice~\cite{Caselle:2025elf,Takahashi:2025plr}.
The entanglement entropy of hadrons has also been shown to be linked to the dynamics of inelastic scattering~\cite{Kharzeev:2017qzs,Tu:2019ouv,Dvali:2021ooc,Hentschinski:2024gaa}, providing phenomenological motivation to studying the \FTEE of baryons.

Finally, an interesting extension of our work is to study \FTEE at finite temperature.
While \FTEE cannot strictly be understood as an entanglement entropy in the deconfined regime (the density matrix represents a mixed state at high temperature), we still observe interesting phenomena in our preliminary results.
Notably, the clean $\ln N_c$ signature seen when a flux tube crosses the boundary in the confined regime disappears in deconfinement; instead, the entropy depends strongly on the entangling region's proximity to the quark or antiquark.
Understanding \FTEE in deconfinement can allow us to study conjectured ``long string states'' near the deconfinement transition~\cite{Kharzeev:2014pha} as well as nonperturbative features that persist above the deconfinement transition.
We plan to report on all of these issues in the future.

\section*{Acknowledgements}
We thank Z. Komargodski for useful comments and discussions. 
R.A. is supported by the Simons Foundation under Award number 994318 (Simons Collaboration on Confinement and QCD
Strings). S.S. and R.A. (partially) are supported by the National Science Foundation under award PHY-2412963. 
In addition, R.A. is supported in part by the Office of Science, Office of Nuclear Physics,
 U.S. Department of Energy under Contract No. DEFG88ER41450 and by the National Science Foundation under award PHY-2412963. R.V. is supported by the U.S. Department of Energy, Office of Science under contract DE-SC0012704. R.V.'s work on quantum information science is supported by the U.S. Department of Energy, Office of Science, National
 Quantum Information Science Research Centers, Co-design Center for Quantum Advantage (C$^2$QA) under contract number  DE-SC0012704. R.V. was also supported at Stony Brook by the Simons Foundation as a co-PI under Award number 994318 (Simons Collaboration on Confinement and QCD Strings). R.V. acknowledges support from the Royal Society Wolfson Foundation Visiting Fellowship and the hospitality of the Higgs Center at the University of Edinburgh. 
The authors thank Stony Brook Research Computing and Cyberinfrastructure and the Institute for Advanced Computational
Science at Stony Brook University for access to the Seawulf HPC system, which was made possible by grants from the
National Science Foundation (awards 1531492 and 2215987) and matching funds from the Empire State Development’s Division
of Science, Technology and Innovation (NYSTAR) program (contract C210148).

\appendix

\section{Representation Flux Basis and (1+1)D \FTEE Derivation}
\label{sec:AppendixFlux}
In this section we discuss the representation flux basis, which provides useful insight into \FTEEformat. Specifically, we will rederive the (1+1)D \FTEE result of $\tilde{S}=F\ln N_c$ we obtained previously in Ref.~\cite{Amorosso:2024glf} in terms of flux states.
In lattice gauge theory, the Hilbert space $\mathcal{H}$ of a group $G$ on a single gauge link is spanned by the basis $\vert U\rangle$ for all $U\in G$.
We can also define an alternative basis~\cite{Lin:2018bud}, related to the former by
\begin{equation}
    \vert R_{\alpha\beta}\rangle\equiv\int\limits_G dU R_{\alpha\beta}(U)\vert U\rangle\,.
\end{equation}
In this expression, $R_{\alpha\beta}(U)$ is the $(\alpha,\beta)$ component of the irreducible representation $R$ of group element $U$ of the group $G$, and the integral with respect to $dU$ represents Haar integration.

The $\vert R_{\alpha\beta}\rangle$ states form the representation flux basis and also span $\mathcal{H}$ for all valid $(R,\alpha,\beta)$ pairings, where $\alpha,\beta\in\{1,\dots, dimR\}$.
The representation flux basis makes constructing gauge-invariant states much easier: If link $l$ is an incoming link to a vertex and link $m$ is an outgoing link from the same vertex, the product state $\vert R_{\alpha\beta}\rangle_l\otimes\vert R_{\beta\gamma}\rangle_m$, where the repeated index $\beta$ is summed over, is gauge-invariant at the shared vertex.\footnote{For one spatial dimension, specifying the representation $R$ of the flux on each gauge link suffices to specify a gauge-invariant state. For higher dimensions, one also needs intertwiners~\cite{Burgio:1999tg}, tensors that track how the $(\alpha,\beta)$ indices of incident links to a vertex combine in a gauge-invariant fashion.}
At a sourceless vertex, gauge invariance requires the representations $R$ of the incoming and outgoing links to match.
In the presence of a source of representation $Q$ at the shared vertex, the picture is more complicated, as the representations of the two links can differ, and one must add a gauge-invariant tensor relating the indices carried by the source, incoming, and outgoing gauge links.

For the sake of completeness, let us also define the operators that can act on these single link states.
We have magnetic operators, as well as link change operators.
Magnetic operators $\hat{R}_{\alpha\beta}$ are defined such that $\hat{R}_{\alpha\beta}\vert U\rangle=R_{\alpha\beta}(U)\vert U\rangle$.
When a magnetic operator acts on the trivial identity representation state $\vert 0\rangle$, we have $\hat{R}_{\alpha\beta}\vert 0\rangle=\vert R_{\alpha\beta}\rangle$.
Link change operators $\hat{L}^{(L/R)}_g$ are associated with a group element $g$ and can act from the incoming or outgoing vertex (right or left in (1+1)D, as seen in the superscript):
$\hat{L}^{(L)}_g\vert U\rangle=\vert gU\rangle$ and $\hat{L}^{(R)}_g\vert U\rangle=\vert Ug\rangle$.

As a final preliminary, we now define the left/right electric field operators $\hat{E}^a_{(L/R)}$ as the generators of left and right link change operators respectively.
Their commutation relations are~\cite{Kogut:1974ag}
\begin{equation}
    [\hat{E}^a_L,\hat{R}_{\alpha\beta}]=R_{\alpha\beta}(T^aU)
\end{equation}
and
\begin{equation}
    [\hat{E}^a_R,\hat{R}_{\alpha\beta}]=R_{\alpha\beta}(UT^a)
\end{equation}
where $T^a$ are the generators of the gauge group.

We can now examine (1+1)D \FTEE through these flux states.
Consider a (1+1)D lattice with infinite $x$-extent.
The Kogut-Susskind Hamiltonian for (1+1)D $SU(N)$ Yang-Mills is~\cite{Kogut:1974ag}
\begin{equation}
    \hat{H}=\frac{1}{2}g^2a\sum\limits_{l}\hat{E}_L\hat{E}_L\,
\end{equation}
where the sum is taken over spatial links $l$. 
The Hamiltonian acting on a single link in state $\vert R_{\alpha\beta}\rangle$ has eigenvalue proportional to the quadratic Casimir of the representation $R$, $C_2(R)$.
This is shown below, taking advantage of the fact that $\hat{E}_L\vert0\rangle=0$~\cite{Kogut:1974ag}:
\begin{align}
    \hat{H}\vert{R_{\alpha\beta}\rangle}&\propto \hat{E}^2_L\vert R_{\alpha\beta}\rangle
    =[\hat{E}^{2}_L,\hat{R}_{\alpha\beta}]\vert0\rangle
    =\hat{E}^{a}_L[\hat{E}^{a}_L,\hat{R}_{\alpha\beta}]\vert0\rangle\nonumber\\
    &=\hat{E}^{a}_LR(T^aU)_{\alpha\beta}\vert0\rangle
    =\hat{E}^{a}_LR(T^a)_{\alpha\gamma}R(U)_{\gamma\beta}\vert0\rangle
    =R(T^a)_{\alpha\gamma}\hat{E}^{a}_L\hat{R}_{\gamma\beta}\vert0\rangle\nonumber\\
    &=R(T^a)_{\alpha\gamma}[\hat{E}^{a}_L,\hat{R}_{\gamma\beta}]\vert0\rangle
    =R(T^a)_{\alpha\gamma}R(T^aU)_{\gamma\beta}\vert0\rangle
    =R(T^aT^aU)_{\alpha\beta}\vert0\rangle\nonumber\\
    &=C_2(R)R(U)_{\alpha\beta}\vert0\rangle
    =C_2(R)\vert R_{\alpha\beta}\rangle\,.
\end{align}
For (1+1)D $SU(N)$ in the absence of sources, the ground state is then the product state of $\vert0\rangle$ on all links, as it will have the minimum eigenvalue (0) of the Hamiltonian.
With fundamental sources, one must have some links carrying non-trivial flux; otherwise, the sole index $\delta$ of the source $Q_\delta$ cannot contract with a gauge-invariant tensor.
As the $x$-extent of the lattice is infinite, the links outside of the two sources must carry trivial flux to minimize the energy.
The minimum energy solution is then the links between the two sources carrying flux in the fundamental representation, with the source indices contracted with the index of the attached link carrying fundamental flux.

Flux tube entanglement entropy, the difference in entanglement entropy in the presence and absence of quarks, is then straightforward to understand.
In the absence of quarks, the ground state can be expressed as a product state, $\Psi=\Psi_V\otimes\Psi_{\bar{V}}$.
For any region $V$ of the lattice, $\Psi_V$ is $\bigotimes\limits_{l\in V}\vert0\rangle_l$ and $\Psi_{\bar{V}}$ is $\bigotimes\limits_{l\in \bar{V}}\vert0\rangle_l$.
As the ground state can be expressed as a product state, the entanglement entropy is zero.
In the presence of quarks, any boundary that partitions this line of fundamental flux between the two quarks brings about entanglement.
For a region $V$ cutting this flux line, we cannot write the ground state as a product state.
Gauge invariance forces indices to be contracted across shared vertices, and every boundary vertex must have its incoming index match its outgoing index, each of which can take $N_c$ values for fundamental-representation flux.
We then have an additional $\ln N_c$ entanglement entropy per boundary crossing in the presence of static quarks, matching our \FTEE result of $F\ln N_c$ from Haar integration in (1+1)D.

\section{Monte Carlo Parameters and Scaling Behavior
  \label{sec:Appendix}}
We present here tables listing the Monte Carlo parameters for our lattice results. We also present scaling plots of the entanglement radius for gauge groups $SU(2)$ through $SU(5)$.
\begin{table}[ht!]
\centering
\caption{\label{tab:lattices_runsSU2}
Monte Carlo simulation parameters for single-slab runs in (2+1)D $SU(2)$ Yang-Mills theory on a $(L_x\times L_y\times L_t)$ lattice with $q$ replicas.
The zero and finite temperature string tensions $a\sqrt{\sigma_0}$ and $a\sqrt{\sigma}$, as well as the width of the slab $w$ and number of top level Polyakov loop correlator samples $N_{\text{samples}}$ are listed as well.
  }
\begin{tabular}
{r|r|r|r|r|r r|r|r|r}
\hline\hline
$q$ & $L_{x}\times L_{y}$ & $L_t$ & $\beta$ 
& $a\sqrt{\sigma}_{0}$ \cite{Teper:1998te} 
& $a\sqrt{\sigma}$ \cite{Amorosso:2024leg} &[$\chi^2$/dof]
&$N_\text{samples}$ 
& $w/a$ & $w\sqrt{\sigma_0}$
\\
\hline
2 & $64\times 16$ & 8 & 6.536 & 0.2307(106) 
& 0.2144(4) & [9.46/1]
& 4362 & 1 & 0.231(11)
\\
2 & $128\times 32$ & 16 & 12.630 & 0.1124(31) 
& 0.1046(2) & [0.22/1]
& 1363 & 2 & 0.225(6)
\\
2 & $192\times 48$ & 24 & 18.679 & 0.0745(17) 
& 0.0695(4) & [0.08/1]
& 571 & 3 & 0.223(5)
\\
2 & $256\times 64$ & 32 & 24.744 & 0.0557(11) 
& 0.0525(2) & [0.05/1]
& 5739 & 4 & 0.223(5)
\\
3 & $256\times 64$ & 32 & 24.744 & 0.0557(11) 
& 0.0525(2) & [0.05/1]
& 1607 & 4 & 0.223(5)
\\
2 & $320\times 80$ & 40 & 30.824 & 0.0444(8) 
& 0.0417(3) & [0.003/1]
& 1165 & 5 & 0.222(4)
\\
2 & $384\times 96$ & 48 & 36.904 & 0.0369(7) 
& 0.0345(3) & [0.17/1]
& 682 & 6 & 0.222(4)
\\\hline\hline
\end{tabular}
\end{table}
\begin{table}[ht!]
\centering
\caption{\label{tab:lattices_runsSU2exotic}
Monte Carlo simulation parameters for double-slab runs in (2+1)D $SU(2)$ Yang-Mills theory on a $(256\times 128\times 64)$ lattice with two replicas.
The width of each slab $w$ and number of top level Polyakov loop correlator samples $N_{\text{samples}}$ are listed as well as parameters $\Delta x$ and $\Delta y$ defined in Sec. \ref{sec:EntanglementRadius}.
  }
\begin{tabular}
{r|r|r|r|r r|r|r|r}
\hline\hline
$\beta$ 
&$N_\text{samples}$ 
& $w\sqrt{\sigma_0}$
& $\Delta x\sqrt{\sigma_0}$
& $\Delta y\sqrt{\sigma_0}$
\\
\hline
24.744 & 1664 &0.223(5) 
& 0 & 0
\\
24.744 & 1535 &0.223(5) 
& 0 & 0.111(2)
\\
24.744 & 2128 &0.223(5) 
& 0 & 0.223(5)
\\
24.744 & 2763 &0.223(5) 
& 0.056(1) & 0.223(5)
\\
24.744 & 3008 &0.223(5) 
& 0.111(2) & 0.223(5)
\\
24.744 & 2562 &0.223(5) 
& 0.167(3) & 0.223(5)
\\
24.744 & 1431 &0.223(5) 
& 0.223(5) & 0.223(5)
\\
24.744 & 1456 &0.223(5) 
& 0.334(7) & 0.223(5)
\\\hline\hline
\end{tabular}
\end{table}
\begin{table}[ht!]
\centering
\caption{\label{tab:lattices_runsSU3}
Monte Carlo simulation parameters for single-slab runs in (2+1)D $SU(3)$ Yang-Mills theory on a $(L_x\times L_y\times L_t)$ lattice with two replicas.
The zero and finite temperature string tensions $a\sqrt{\sigma_0}$ and $a\sqrt{\sigma}$, as well as the width of the slab $w$ and number of top level Polyakov loop correlator samples $N_{\text{samples}}$ are listed as well.
}
\begin{tabular}
{r|r|r|r|r r|r|r|r}
\hline\hline
$L_{x}\times L_{y}$ & $L_t$ & $\beta$ 
& $a\sqrt{\sigma}_{0}$ \cite{Teper:1998te} 
& $a\sqrt{\sigma}$ &[$\chi^2$/dof]
&$N_\text{samples}$ 
& $w/a$ & $w\sqrt{\sigma_0}$
\\
\hline
$64\times 32$ & 8 & 15.329 & 0.2492(30) 
& 0.2482(22) & [11.27/2]
& 561 & 1 & 0.249(3)
\\
$128\times 64$ & 16 & 28.542 & 0.1250(11) 
& 0.1240(7) & [0.74/2]
& 665 & 2 & 0.250(2)
\\
$192\times 96$ & 24 & 41.896 & 0.0829(7) 
& 0.0822(4) & [0.13/2]
& 441 & 3 & 0.249(2)
\\
$256\times 128$ & 32 & 55.302 & 0.0620(5) 
& 0.0618(4) & [0.06/2]
& 745 & 4 & 0.248(2)
\\
$320\times 160$ & 40 & 68.730 & 0.0494(4) 
& 0.0487(62) & [4.34/2]
& 323 & 5 & 0.247(2)
\\\hline\hline
\end{tabular}
\end{table}
\begin{table}[ht!]
\centering
\caption{\label{tab:lattices_runsSU4}
Monte Carlo simulation parameters for single-slab runs in (2+1)D $SU(4)$ Yang-Mills theory on a $(L_x\times L_y\times L_t)$ lattice with two replicas.
The zero and finite temperature string tensions $a\sqrt{\sigma_0}$ and $a\sqrt{\sigma}$, as well as the width of the slab $w$ and number of top level Polyakov loop correlator samples $N_{\text{samples}}$ are listed as well.
  }
\begin{tabular}
{r|r|r|r|r r|r|r|r}
\hline\hline
$L_{x}\times L_{y}$ & $L_t$ & $\beta$ 
& $a\sqrt{\sigma}_{0}$ \cite{Teper:1998te} 
& $a\sqrt{\sigma}$ &[$\chi^2$/dof]
&$N_\text{samples}$ 
& $w/a$ & $w\sqrt{\sigma_0}$
\\
\hline
$128\times 64$ & 16 & 51.000 & 0.1284(2) 
& 0.1282(6) & [0.35/2]
& 593 & 2 & 0.257(0)
\\
$192\times 96$ & 24 & 74.000 & 0.0864(1) 
& 0.0862(3) & [0.02/2]
& 392 & 3 & 0.259(0)
\\
$256\times 128$ & 32 & 98.976 & 0.0637(4) 
& 0.0642(17) & [2.97/2]
& 484 & 4 & 0.255(2)
\\
$320\times 160$ & 40 & 121.237 & 0.0516(3) 
& 0.0499(19) & [3.38/2]
& 152 & 5 & 0.258(2)
\\\hline\hline
\end{tabular}
\end{table}
\begin{table}[ht!]
\centering
\caption{\label{tab:lattices_runsSU5}
Monte Carlo simulation parameters for single-slab runs in (2+1)D $SU(5)$ Yang-Mills theory on a $(L_x\times L_y\times L_t)$ lattice with two replicas.
The zero and finite temperature string tensions $a\sqrt{\sigma_0}$ and $a\sqrt{\sigma}$, as well as the width of the slab $w$ and number of top level Polyakov loop correlator samples $N_{\text{samples}}$ are listed as well.
  }
\begin{tabular}
{r|r|r|r|r r|r|r|r}
\hline\hline
$L_{x}\times L_{y}$ & $L_t$ & $\beta$ 
& $a\sqrt{\sigma}_{0}$ \cite{Teper:1998te} 
& $a\sqrt{\sigma}$ &[$\chi^2$/dof]
&$N_\text{samples}$ 
& $w/a$ & $w\sqrt{\sigma_0}$
\\
\hline
$128\times 64$ & 16 & 84.000 & 0.1272(3) 
& 0.1229(10) & [0.58/2]
& 597 & 2 & 0.254(1)
\\
$192\times 96$ & 24 & 115.155 & 0.0883(6) 
& 0.0880(2) & [1.28/2]
& 89 & 3 & 0.265(2)
\\
$256\times 128$ & 32 & 154.650 & 0.0649(4) 
& 0.0644(6) & [0.20/2]
& 287 & 4 & 0.260(2)
\\
$320\times 160$ & 40 & 191.948 & 0.0519(3) 
& 0.0513(2) & [0.07/1]
& 56 & 5 & 0.260(2)
\\\hline\hline
\end{tabular}
\end{table}
\begin{figure}[ht!]
\centering
\includegraphics[width=.49\textwidth]{iwscaling.pdf}
\includegraphics[width=.49\textwidth]{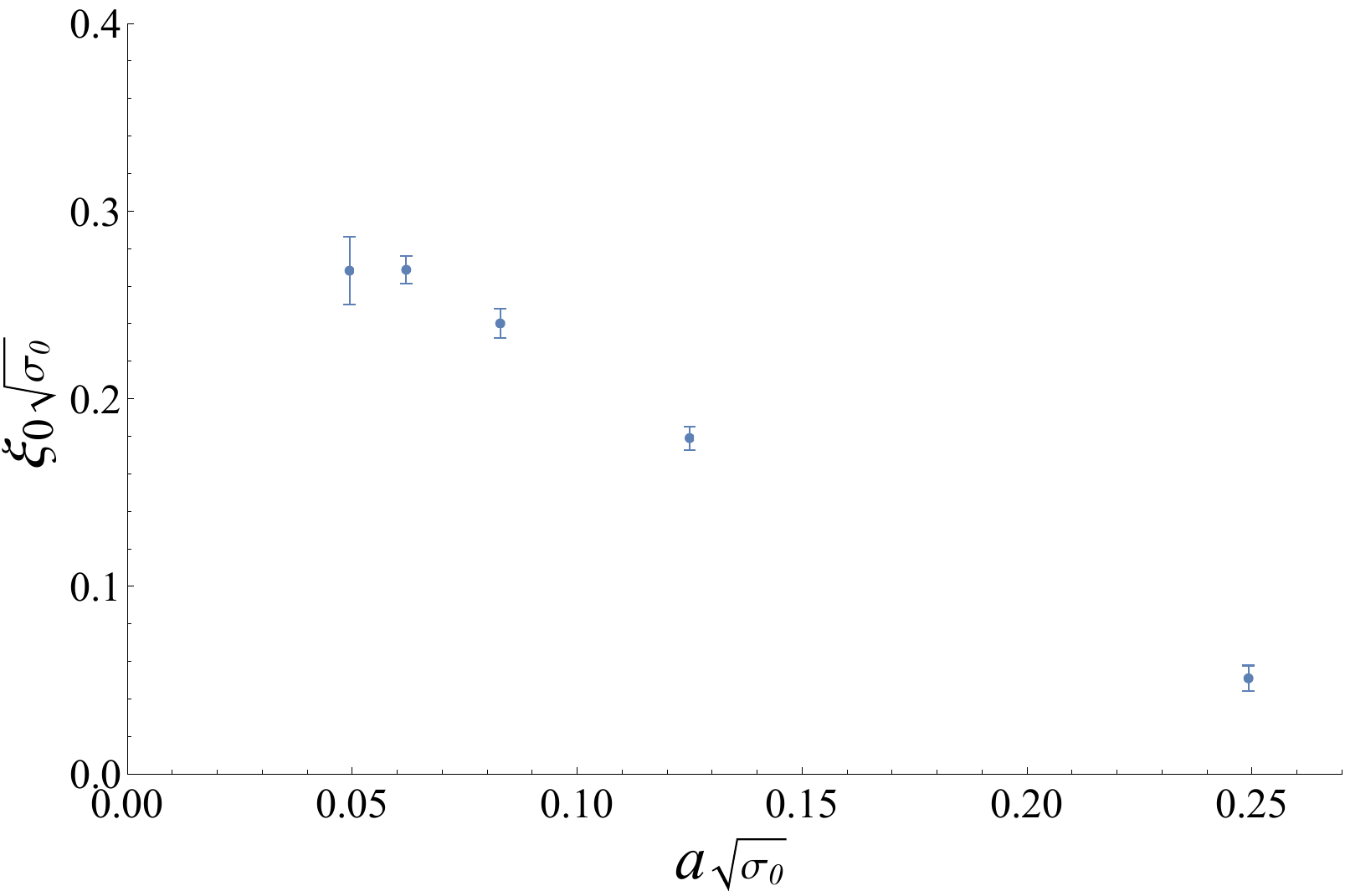}
\caption{\label{fig:iwscaling}
Scaling plot of the entanglement radius for $SU(2)$ (left) and $SU(3)$ (right) (2+1)D Yang-Mills theory. 
The entanglement radius for both gauge groups is determined using the offset method with parameters outlined in Tables \ref{tab:lattices_runsSU2} and \ref{tab:lattices_runsSU3}.
}
\end{figure}
\begin{figure}[ht!]
\centering
\includegraphics[width=.49\textwidth]{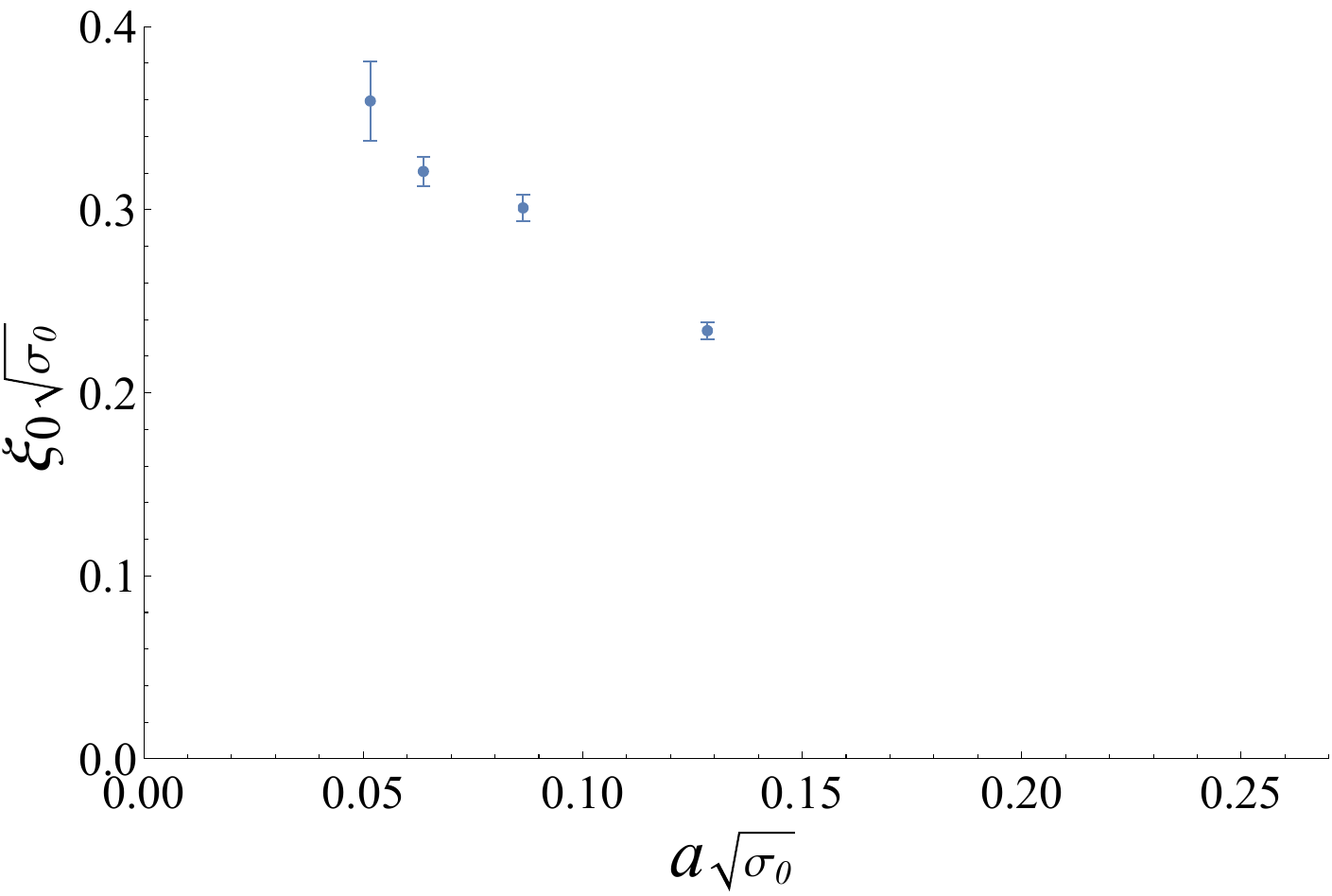}
\includegraphics[width=.49\textwidth]{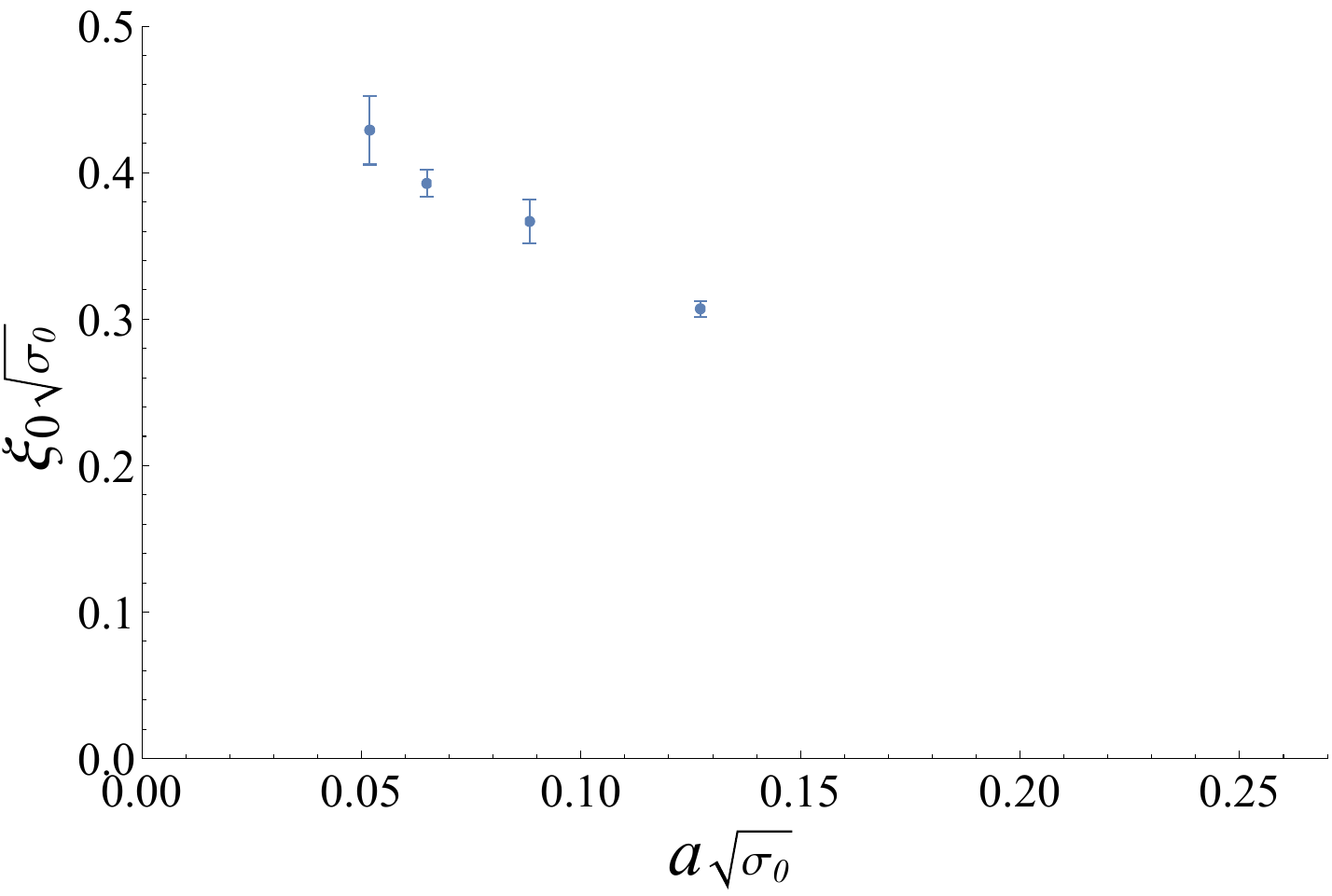}
\caption{\label{fig:iwscalinglargeN}
Scaling plot of the entanglement radius for $SU(4)$ (left) and $SU(5)$ (right) (2+1)D Yang-Mills theory. 
The entanglement radius for both gauge groups is determined using the offset method with parameters outlined in Tables \ref{tab:lattices_runsSU4} and \ref{tab:lattices_runsSU5}.
}
\end{figure}
\FloatBarrier
\bibliography{bib}

\end{document}